\documentclass[twocolumn]{aastex702}

\usepackage{ragged2e}
\usepackage{amsmath}
\usepackage[nameinlink]{cleveref}
\usepackage[]{natbib}
\setcitestyle{comma}
\usepackage{array}
\usepackage{longtable}
\usepackage{tabularx}
\usepackage{multirow}
\usepackage[para]{threeparttable}
\usepackage{makecell}
\usepackage{booktabs}
\usepackage[caption=false]{subfig}

\newcolumntype{Y}[1]{>{\centering\arraybackslash}m{#1}}

\usepackage{comment}

\usepackage[utf8]{inputenc}
\usepackage[T1]{fontenc}

\makeatletter
\newcommand*{\linktocite}[2]{%
  \hyper@natlinkstart{#1}#2\hyper@natlinkend}
\makeatother

\newcommand{\Eq}[1]{Equation\,(\ref{#1})}

\newcommand{\Sec}[1]{Section~\ref{#1}}

\newcommand{\Fig}[1]{Figure~\ref{#1}}

\def\K{\; {\rm K}}

\begin{document}
\title{The climates and thermal emission spectra of prime nearby temperate rocky exoplanet targets}

\correspondingauthor{Tobi Hammond}
\email{tobi.hammond12@gmail.com}
\author[0009-0002-5756-9778]{Tobi Hammond}
\affiliation{Department of Earth, Atmospheric, and Planetary Science, Purdue University, West Lafayette, IN, 47907, USA}
\affiliation{NASA Goddard Space Flight Center, Greenbelt, MD 20771, USA}
\affiliation{Center for Research and Exploration in Space Science and Technology, NASA Goddard Space Flight Center, Greenbelt, MD, USA}
\affiliation{Department of Astronomy, University of Maryland, College Park, MD 20742, USA}
\email{fakeemail1@google.com}  

\author[0000-0002-9258-5311]{Thaddeus D. Komacek}
\affiliation{Department of Physics (Atmospheric, Oceanic and Planetary Physics), University of Oxford, Oxford OX1 3PU, UK}
\affiliation{Department of Astronomy, University of Maryland, College Park, MD 20742, USA}
\affiliation{Blue Marble Space Institute of Science, Seattle, WA 98104, USA}
\email{fakeemail2@google.com}  

\author[0000-0002-5893-2471]{Ravi K Kopparapu}
\affiliation{NASA Goddard Space Flight Center, Greenbelt, MD 20771, USA}
\affiliation{CHAMPs Collaboration}
\email{fakeemail3@google.com}  

\author[0000-0002-5967-9631]{Thomas J Fauchez}
\affiliation{NASA Goddard Space Flight Center, Greenbelt, MD 20771, USA}
\affiliation{American University, 4400 Massachusetts Avenue, NW, Washington, DC, 20016, USA}
\email{fakeemail4@google.com}  

\author[0000-0002-8119-3355]{Avi M Mandell}
\affiliation{NASA Goddard Space Flight Center, Greenbelt, MD 20771, USA}
\email{fakeemail5@google.com}  

\author[0000-0002-7188-1648]{Eric T Wolf}
\affiliation{Laboratory for Atmospheric and Space Physics, University of Colorado Boulder, Boulder, CO, USA}
\affiliation{Blue Marble Space Institute of Science, Seattle, WA, USA}
\email{fakeemail6@google.com}  

\author[0000-0002-5060-1993]{Vincent Kofman}
\affiliation{NASA Goddard Space Flight Center, Greenbelt, MD 20771, USA}
\affiliation{American University, 4400 Massachusetts Avenue, NW, Washington, DC, 20016, USA}
\email{fakeemail7@google.com}  

\author[0000-0002-7084-0529]{Stephen R Kane}
\affiliation{Department of Earth and Planetary Sciences, University of California, Riverside, CA 92521, USA}
\email{fakeemail8@google.com}  

\author[0000-0002-1570-2203]{Ted M Johnson}
\affiliation{Nevada Center for Astrophysics, University of Nevada, Las Vegas, 4505 South Maryland Parkway, Las Vegas, NV 89154, USA}
\affiliation{Department of Physics and Astronomy, University of Nevada, Las Vegas, 4505 South Maryland Parkway, Las Vegas, NV 89154, USA}
\affiliation{NASA Goddard Space Flight Center, Greenbelt, MD 20771, USA}
\email{fakeemail9@google.com}  

\author[0009-0000-9206-5589]{Anmol Desai}
\affiliation{NASA Goddard Space Flight Center, Greenbelt, MD 20771, USA}
\affiliation{Center for Research and Exploration in Space Science and Technology, NASA Goddard Space Flight Center, Greenbelt, MD, USA}
\affiliation{Southeastern Universities Research Association, Washington, DC, USA}
\email{fakeemail10@google.com}  

\author[0000-0001-6285-267X]{Giada Arney}
\affiliation{NASA Goddard Space Flight Center, Greenbelt, MD 20771, USA}
\email{fakeemail11@google.com}  

\author[0000-0003-2273-8324]{Jaime S. Crouse}
\affiliation{NASA Goddard Space Flight Center, Greenbelt, MD 20771, USA}
\affiliation{Johns Hopkins University, Baltimore, MD 21218}
\email{fakeemail12@google.com}  


\begin{abstract}

Over the course of the past decade, advances in the radial velocity and transit techniques have enabled the detection of rocky exoplanets in the habitable zones of nearby stars. Future observations with novel methods are required to characterize this sample of planets, especially those that are non-transiting. One proposed method is the Planetary Infrared Excess (PIE) technique, which would enable the characterization of non-transiting planets by measuring the excess infrared flux from the planet relative to the star's spectral energy distribution. In this work, we predict the efficacy of future observations using the PIE technique by potential future observatories such as the MIRECLE mission concept. To do so, we conduct a broad suite of 21 General Circulation Model (GCM) simulations with ExoCAM of seven nearby habitable zone targets for three choices of atmospheric composition with varying partial pressure of CO$_2$. We then construct thermal phase curves and emission spectra by post-processing our ExoCAM GCM simulations with the Planetary Spectrum Generator (PSG). We find that all cases have distinguishable carbon dioxide and water features assuming a 90$^\circ$ orbital inclination. Notably, we predict that CO$_2$ is potentially detectable with MIRECLE for at least four nearby known non-transiting rocky planet candidate targets in the habitable zone: Proxima Cenaturi b, GJ 1061 d, GJ 1002 b, and Teegarden's Star c. Our ExoCAM GCMs and PSG post-processing demonstrate the potential to observationally characterize nearby non-transiting rocky planets and better constrain the potential for habitability in our Solar neighborhood.
\end{abstract}

\section{Introduction}

There has been a rapid advance in the discovery space of potentially habitable exoplanets over the past decade, including the discoveries of Proxima Centauri b \citep{Anglada-Escude:2016aa}, TRAPPIST-1e,f,g \citep{Gillon:2017aa}, TOI-700d,e \citep{2020AJ....160..117R,2023arXiv230103617G}, Wolf 1069 b \citep{Kossakowski_2023}, and LP 890-9c \citep{2022A&A...667A..59D}, among others. These detections are opening up a new era of exoplanet science that began with the commissioning of JWST, enabling infrared spectral characterization to constrain the presence or absence of atmospheres on rocky exoplanets \citep{2023arXiv230104191L,Greene:2023aa,May:2023aa,Zieba:2023aa,Mansfield:2024aa,Xue:2024aa}. JWST is expected to be able to probe the atmospheric composition of nearby transiting \citep{Lustig-Yaeger:2019aa,Mikal-Evans:2022uk} and possibly non-transiting exoplanets \citep{Kreidberg:2016aa,2020ApJ...898L..35S} orbiting M dwarf stars, though stellar activity can be a significant challenge \citep{Lim:2023aa,Rackham:2024aa}.  Additionally, future mission concepts targeting UV/optical/NIR imaging such as the Habitable Worlds Observatory as well as mid-IR spectroscopy and phase curves with missions such as MIRECLE \citep{Mandell_2022} and the Far-IR Surveyor and LIFE \citep{2022A&A...664A..21Q} flagships would further advance our understanding of rocky exoplanets by enabling characterization of their atmospheres.

Many of the nearby rocky planets in the habitable zone are non-transiting and have close-in orbits around M-dwarf host stars, because our current detection methods favor finding Earth-sized planets around smaller and less massive stars \citep{Lovis_2009,Winn_2009}. Characterizing these planets requires the development of alternate methods beyond primary and secondary eclipse spectroscopy or traditional direct imaging to observationally characterize their atmospheres. A clear path forward for ground-based observatories is the combination of high dispersion spectroscopy with high contrast imaging (HDS + HCI, \citealp{Snellen:2015aa,Vaughan:2024aa}) that can theoretically reach contrast limits of $10^{-10}$. 

One recently developed method for space-based observatories is the Planetary Infrared Excess, or PIE, technique \citep{2020ApJ...898L..35S}. The PIE technique targets the planetary infrared excess (as the name suggests) and its presence in the unresolved combined stellar and planetary spectra. PIE attempts to isolate the fraction of light originating from the planet by accurately modeling the stellar spectral energy distribution (SED). PIE is similar to the detection of debris disks via infrared excess in that it separates the spectra of the planet and star in order to isolate the planetary thermal emission. The PIE technique has been demonstrated to be feasible for simulated JWST observations of hot Jupiters \citep{Lustig-Yaeger:2021aa} as well as simulated observations of Proxima Centauri \citep{Mandell_2022} and the TRAPPIST-1 system \citep{Mayorga:2023aa}. Using a similar technique but with higher spectral resolution, \cite{Snellen:2017} proposed that high-pass spectral filtering of planetary thermal emission could enable detection of the $15~\mu\mathrm{m}$ CO$_2$ band with MIRI MRS. 

The PIE technique has the best potential to constrain planetary climate when measuring the direct planetary thermal emission over many orbital phases. As a result, PIE is strongly dependent on the spatial pattern of thermal flux from the planet as it orbits its host star. However, to date there have been no multi-dimensional model predictions for the PIE technique. The PIE technique is further best suited to characterize nearby planets orbiting small, cool stars, given the requirement that the planetary flux at long wavelengths must be above the noise floor \citep{2020ApJ...898L..35S}. Studies with three-dimensional (3D) General Circulation Models (GCMs) of temperate rocky planets in the habitable zones of small K and M dwarf systems have found that their temperature and cloud patterns are strongly inhomogeneous, leading to a strong phase dependence in resulting thermal emission  \citep{Yang:2013,Koll:2016,Turbet:2016aa,Wolf:2017aa,way:2018,Shields:2019aa,Batra:2024aa,Lobo:2024aa}. A range of previous GCM simulations have been applied to study the potential climate dynamics of nearby temperate rocky planets, including Proxima Centauri b \citep{Turbet:2016aa,Boutle:2017aa,Genio:2017aa,Lewis:2018aa,Salazar:2020aa,DeLuca:2024aa}, TRAPPIST-1e \citep{Wolf:2017aa,2018A&A...612A..86T,Fauchez:2019aa,May:2021aa,2022PSJ.....3..211T,2022PSJ.....3..212S,2022PSJ.....3..213F,Rotman:2023aa,Mak:2024aa}, TRAPPIST-1d \citep{Wolf:2017aa,Turbet:2023aa}, and TOI-700d \citep{Suissa:2020b}. Though there have been a range of generalized studies of temperate rocky planets orbiting M dwarf stars with varying planetary parameters (e.g., \citealp{Joshi:1997,Merlis:2010,Yang:2014,Koll:2016,Fujii:2017aa,Shields:2019aa,Wolf:2019aa,Suissa:2020aa,Macondald:2022aa,Turbet:2023aa}), there has not yet been a broad and uniform study of the climates of the range of specific nearby habitable zone targets that be characterized in thermal emission with PIE. 

In this work, we conduct a suite of GCM simulations of seven nearby (1.3 - 32.4 pc) temperate, potentially rocky planets orbiting late-type stars: LP 890-9c, TRAPPIST-1e, GJ 1002 b, Proxima Centauri b, Wolf 1069 b, GJ 1061 d, and Teegarden's Star c. In order to study their broad range of potential climate states, we vary the primary greenhouse gas CO$_2$ over a wide range for each planet case from approximately Earth-like to a thick 2 bar CO$_2$ atmosphere. We post-process our GCM simulations to predict thermal emission spectra and the detectability of spectral features with the PIE technique.

This manuscript is organized as follows. We first describe both our ExoCAM GCM and NASA PSG model setup in \Sec{sec:methods}. We then present our simulated climate and observable properties of seven high-priority nearby rocky planet targets in \Sec{sec:results}. We discuss our findings, limitations, and future work in \Sec{sec:disc}, and state key takeaways in \Sec{sec:conc}. 

\section{Methods}
\label{sec:methods}
\subsection{GCM Model}
To simulate the atmospheres of the selected seven synchronously rotating rocky exoplanets orbiting M-dwarf stars, we used ExoCAM\footnote{https://github.com/storyofthewolf/ExoCAM}, a 3D general circulation model \citep{Wolf_2022}. ExoCAM has been used to investigate the climates of exoplanets across a broad range of parameter space \citep{kopparapu2017,Wolf:2017aa,Haqq2018,Yang:2019aa,Suissa:2020aa,Wei:2020aa,Wolf:2020aa,Wolf:2022aa,Rotman:2023aa,Zhan:2024aa,Hammond:2024aa,Garcia:2024aa} and various model inter-comparison studies \citep{2022PSJ.....3..211T,2022PSJ.....3..212S,2022PSJ.....3..213F}. ExoCAM is a publicly available modified version of the Community Earth System Model (CESM) version 1.2.1 \citep{Neale:2010aa}. ExoCAM is coupled to ExoRT\footnote{https://github.com/storyofthewolf/ExoRT}, a flexible two-stream correlated-k radiative transfer scheme.

We conducted 21 ExoCAM GCMs for 7 planet targets (Table \ref{model_params}), with 3 atmospheric scenarios for each planet with pCO$_2$ varying from $100~\mu\mathrm{bar}$ to $2~\mathrm{bar}$ (Table \ref{tab:atm}). 
\begin{table*}
\resizebox{\textwidth}{!}{
    \begin{threeparttable}
    \caption{Planetary parameters adopted for our GCM simulations. References for these parameters are shown below.}
    \label{model_params}
    \centering
    \begin{tabular}{ccccccc} \hline\hline
         \textbf{Planet Name}&  \textbf{Radius ($R_\oplus$)}&  \textbf{g (${ms^{-2}}$)}&  \textbf{Period (\textit{days})} &\textbf{Instellation (${S_\oplus}$)}& \textbf{Stellar Temp (K)}& \textbf{Distance (\textit{pc)}}\\ \hline \hline
         LP 890-9c\tnote{1}& 1.37& 13.10& 8.46& 0.91& 2871& 32.43 \\ \hline 
         TRAPPIST-1e\tnote{2}& 0.92& 8.01& 6.10 & 0.646& 2566& 12.47\\ \hline 
         GJ 1002 b\tnote{3}& 1.10& 8.74& 10.35& 0.67& 3024& 4.85\\ \hline
         Proxima Centauri b\tnote{4}& 1.10& 10.88& 11.20& 0.645\tnote{5}& 3050& 1.30 \\ \hline 
         Wolf 1069 b\tnote{6}& 1.08& 10.58& 5.60& 0.652& 3158& 9.58\\ \hline 
         GJ 1061 d\tnote{7}& 1.23& 10.61\tnote{8}& 13.03& 0.60& 2953& 3.67\\ \hline 
         Teegarden's Star c\tnote{9}& 1.05& 9.86\tnote{8}& 11.41& 0.37& 2904& 3.83\\ \hline     
    \end{tabular}
    \begin{tablenotes}
        \centering
        \item [1] \cite{Delrez_2022}
        \item [2] \cite{Agol_2021}
        \item [3] \cite{Mascareno_2023}
        \item [4] \cite{Turbet_2016}
        \item [5] \cite{Boutle_2017}
        \item [6] \cite{Kossakowski_2023}
        \item [7] \cite{Dreizler_2020}
        \item [8] \cite {Boldog_2024}
        \item [9] \cite{Zechmeister_2019}
     \end{tablenotes}
    \end{threeparttable}
    }
\end{table*}
\begin{table}[]
\centering
       \caption{Parameter choices for our suite of ExoCAM GCM simulations. We conduct 3 simulations for each planet with varying pCO$_2$, for a total of 21 GCM simulations.}
     \label{tab:atm}
    \begin{tabular}{cccc}
    \hline\hline
    \textbf{CO2 (bar)}& \textbf{N$_2$ (bar)}& \multicolumn{1}{l}{\textbf{Obliquity ($^\circ$})} & \multicolumn{1}{l}{\textbf{Eccentricity}} \\ \hline\hline
    0.0001 & 1 & 0 & 0 \\ \hline
    0.1 & 1 & 0 & 0 \\ \hline
    2 & 0 & 0 & 0 \\ \hline
    \end{tabular}
\end{table}
We choose this set of planet targets in order to both benchmark against previous studies (for e.g., TRAPPIST-1e, Proxima Centauri b) as well as conduct simulations of planets that are expected to be promising targets for the PIE technique. For the non-transiting planets in the sample, for simplicity we take the mass to be the minimum mass from radial velocity observations and we use the radius calculated assuming an Earth-like bulk composition (see Table \ref{model_params} for the sources of these radial velocity masses as well as our other assumed planetary parameters). 

All of our 21 ExoCAM GCMs make the same set of the following key assumptions. First, we assume that each planet is spin-synchronized, with a rotation period equal to its orbital period. We further assume that each planet has zero obliquity and zero eccentricity, consistent with tidal locking. We note that because each planet in the sample are cool and expected to have a ratio of radiative to rotational timescales $\tau_\mathrm{rad}/P_\mathrm{rot} > 1$, the effects of seasonality on climate would be minimal regardless \citep{Ohno_2019_1}. We assume that each target is an aquaplanet with a global surface slab ocean with a depth of 50 m and zero ocean heat transport. We allow sea ice to form thermodynamically \citep{Bitz:2012aa}, but we do not include sea ice drift \citep{Yang:2023aa}. 

Our simulations have a horizontal resolution of  4$^\circ$ x 5$^\circ$ and 40 atmospheric levels. They have a 30-minute physics time step and ratio of physics to dynamical timestep, termed ``nsplit'' in CESM, of 32 (leading to a dynamical timestep of 56.25 seconds) and a 60-minute radiative time step. Each simulation was run until equilibrium in both top-of-atmosphere net radiation and the globally averaged surface temperature. In the following, we only show results from the average of the last ten years of model output. 

\subsection{Planetary Spectrum Generator}
We use the publicly available NASA Planetary Spectrum Generator (PSG)\footnote{\url{https://psg.gsfc.nasa.gov/},\url{https://github.com/nasapsg}} \citep{Villanueva_2018,2022fpsg.book.....V}, specifically its Global Emission Spectra (GlobES) module\footnote{\url{https://psg.gsfc.nasa.gov/apps/globes.php}} \citep{Kofman:2024aa} to simulate thermal emission spectra from our ExoCAM outputs. 
To simulate idealized PIE observations with PSG, we divide the planetary thermal emission by the total flux of the unresolved system, yielding the planet-star contrast. This technique assumes complete knowledge of the stellar spectrum -- i.e., that it can be removed from the total spectrum without resulting in biases in the planetary spectrum. We also assume that the star is noiseless in the initial PIE calculation for simplicity. To relax these assumptions would require more complicated retrieval methods (Johnson et al., in prep)

In order to conduct our PSG GlobES post-processing calculations, we take the ExoCAM output atmospheric composition profiles for H$_2$O, N$_2$, and CO$_2$ at each longitude and latitude as well as the liquid water cloud and ice cloud profiles. Our PSG GlobES calculations take in the 3D GCM output as input and then computes the thermal emission spectra for a specific orbital phase (or viewing geometry) as in \cite{Johnson:2024aa}.
We include Rayleigh scattering, atmospheric refraction, and collision-induced absorption in our calculations, which are all on by default in PSG GlobES. Relevant in particular for this study are the H$_2$O-H$_2$O collision induced absorption, which are parameterized from the MT$\textunderscore$CKD continuum \citep{Kofman:2021aa} and adopted both in PSG and ExoCAM.

\begin{table*}
    \centering
    \caption{Instrumental parameter choices for our PSG radiative transfer post-processing calculations.}
    \begin{tabular}{lc|lc} \hline \hline
    \textbf{Telescope Parameter}& \textbf{Value}& \textbf{Noise Parameter}& \textbf{Value}\\ \hline \hline
         Telescope Diameter (m)&  1.5&  Number of Pixels&  8\\ \hline 
         Field of View (arcsec)&  6 &  Read Noise (e$^{-}$)&   16.8 \\ \hline 
         Resolving power (RP) &  50&  Dark Rate (e$^{-}$/s$^{-1}$)&  100 
        \\ \hline 
        Wavelength Coverage ($\mu$m)& 5 - 18 & Optics temperature (K)& 35 \\ \hline 
        Exposure Time (sec) & 60 & Optics Emissivity&0.1\\ \hline 
         \multicolumn{2}{c|}{}& Total Throughput&0.7\\ \hline
    \end{tabular}
    \label{tab:mirecleparams}
\end{table*}

We calculate thermal emission spectra in the mid-infrared from $5-18~\mu\mathrm{m}$ for a MIRECLE-like telescope with a diameter of $1.5~\mathrm{m}$. Telescope parameters as well as assumed noise parameters are listed in Table \ref{tab:mirecleparams} and are the default settings for MIRECLE in PSG \citep{Mandell_2022}. To provide accurate reflection of our ExoCAM models into PSG, we set the spatial binning value to 3 in order to minimize computation time with minimal effect on the simulated observations. 
We assumed an integration time of 30 days for all noise calculations, taking into account phase variations over multiple orbital phases of each target in our PIE calculations with PSG/GlobES. 

\subsection{Model limitations}
The simulations presented in this work are idealized in order to explore the possible parameter space of nearby rocky planet targets and guide potential observational characterization. Notably, in our GCMs we assume only three possible atmospheric compositions for each planet, all of which presume that N$_2$ or CO$_2$ is the dominant gas. All simulations further assume tidal locking, with a 1:1 spin-orbit ratio. Additionally, note that the radii we choose for non-transiting planets assume a broadly Earth-like interior composition, even though compression will increase the bulk density of super-Earth mass planets \citep{Valencia06}. 

In addition, our post-processed observations with PSG use time-averaged GCM output rather than snapshots to predict the observable spectra and phase curves, but previous work has shown that time variability may impact spectra \citep{2021FrASS...8..134S,May:2021aa}. As a result, spectra at a specific orbital phase could differ from orbit to orbit, which is neglected here. We also limit our phase range for the post-processing, rather than simulating the spectra over a full orbit. 

These idealized setup of our GCMs and radiative transfer post-processing should be kept in mind when viewing the results from these simulations in the following sections. We discuss these and other limitations of the study in greater detail in \Sec{sec:limitations}. 

\section{Results}
\label{sec:results}
\subsection{ExoCAM GCMs}
\subsubsection{Planetary climate}
\begin{figure*}[h]
\begin{minipage}[c]{0.49\linewidth}
\includegraphics[width=\linewidth]{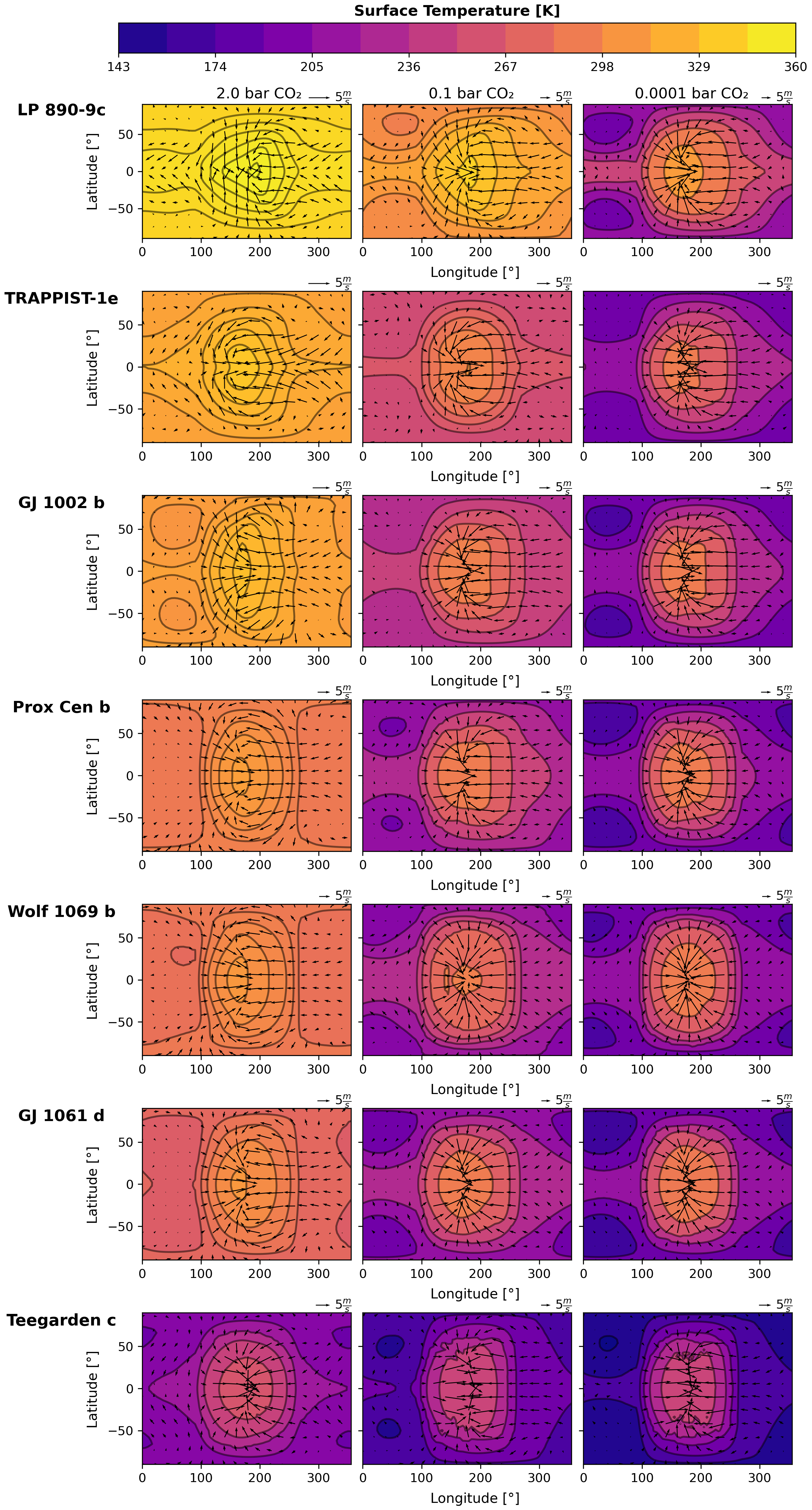}
\caption{Maps of surface temperature (colors) for each GCM case for each of our seven considered targets (rows) and varying pCO$_2$ (columns). Over-plotted are the near-surface winds (quivers). Maps are shown as a function of longitude in degrees (x-axis) and latitude in degrees (y-axis), with the substellar point centered at ($0^\circ,0^\circ$). All cases share the same colorbar for inter-comparison. Lighter colors (yellow) represent higher temperatures, while darker colors (blue) represent lower temperatures. As expected, we find that higher pCO$_2$ leads to warmer surfaces. All planet cases except Teegarden's Star c have some habitable surface area with temperatures above the freezing point of water, while Teegarden's Star c is ice-covered for all pCO$_2$ values considered.}
\label{fig:st_plots}
\end{minipage}
\begin{minipage}[c]{0.49\linewidth}
\includegraphics[width=\linewidth]{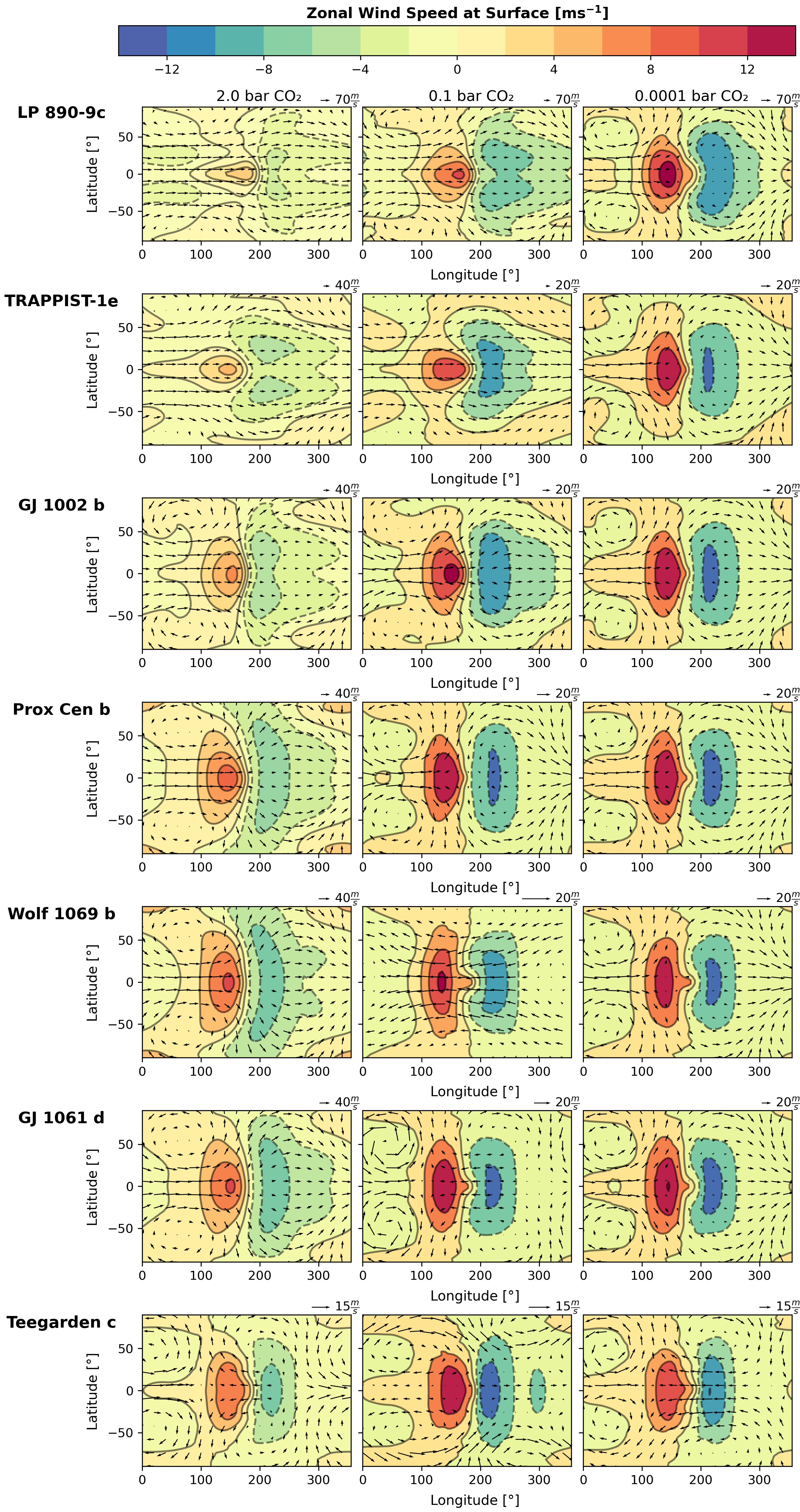}
\caption{Maps of normalized near surface zonal wind speed (colors) with directional wind vectors at 250 mbar (quivers) for each GCM case. The quiver magnitudes are normalized separately for each plot. Red shading represents positive zonal wind speeds, while blue represents negative zonal wind speeds. All cases considered have a superrotating equatorial jet with a maximum wind speed near the western terminator. We find that wind speeds are generally faster for cases with lower pCO$_2$, driven by the larger day-to-night temperature contrasts \citep{Pierrehumbert:2019vk} in our cooler cases. However, winds aloft are generally faster in cases with higher instellation and higher pCO$_2$, with faster superrotating jets aloft. }
\label{fig:wind_plots}
\end{minipage}%
\end{figure*}

We first present the simulated climate dynamics from our 21 total cases of seven different nearby rocky planets.
\Fig{fig:st_plots} shows surface temperature maps (both in filled and open contours) for each of the seven targets (rows) for varying pCO$_2$ (columns). Note also that we provide a table summarizing the global-mean GCM output in Appendix \ref{ref:app}. The targets are listed in order of decreasing instellation from top to bottom, and all sub-plots share a color scheme for inter-comparison. 
As expected, we find that for a given pCO$_2$ the peak dayside temperature and minimum nightside temperature increase with increasing instellation. We also find that the habitability of our cases depends on pCO$_2$, with higher pCO$_2$ leading to warmer surfaces due to the greenhouse effect of CO$_2$ and H$_2$O \citep{Manabe:1975aa}. For instance, in the cases of LP 890-9c, TRAPPIST-1e, and GJ 1002 b, the maximum dayside temperature is above the $50^\circ~\mathrm{C}$ conservative threshold for habitability \citep{Shields:2016aa,Lobo:2023aa,Lobo:2024aa} in high pCO$_2$ cases but below this limit with lower values of pCO$_2$. However, we find that for all the pCO$_2$ cases considered, each of the seven temperate rocky planets considered retains a stable climate below the runaway greenhouse limit. The only exception to our targets having habitable conditions somewhere in our range of considered pCO$_2$ is the case of Teegarden's Star c, which has temperatures below the freezing point of water everywhere on its surface for all atmospheric compositions that we considered. 

\begin{figure}[h]
\includegraphics[width=\linewidth]{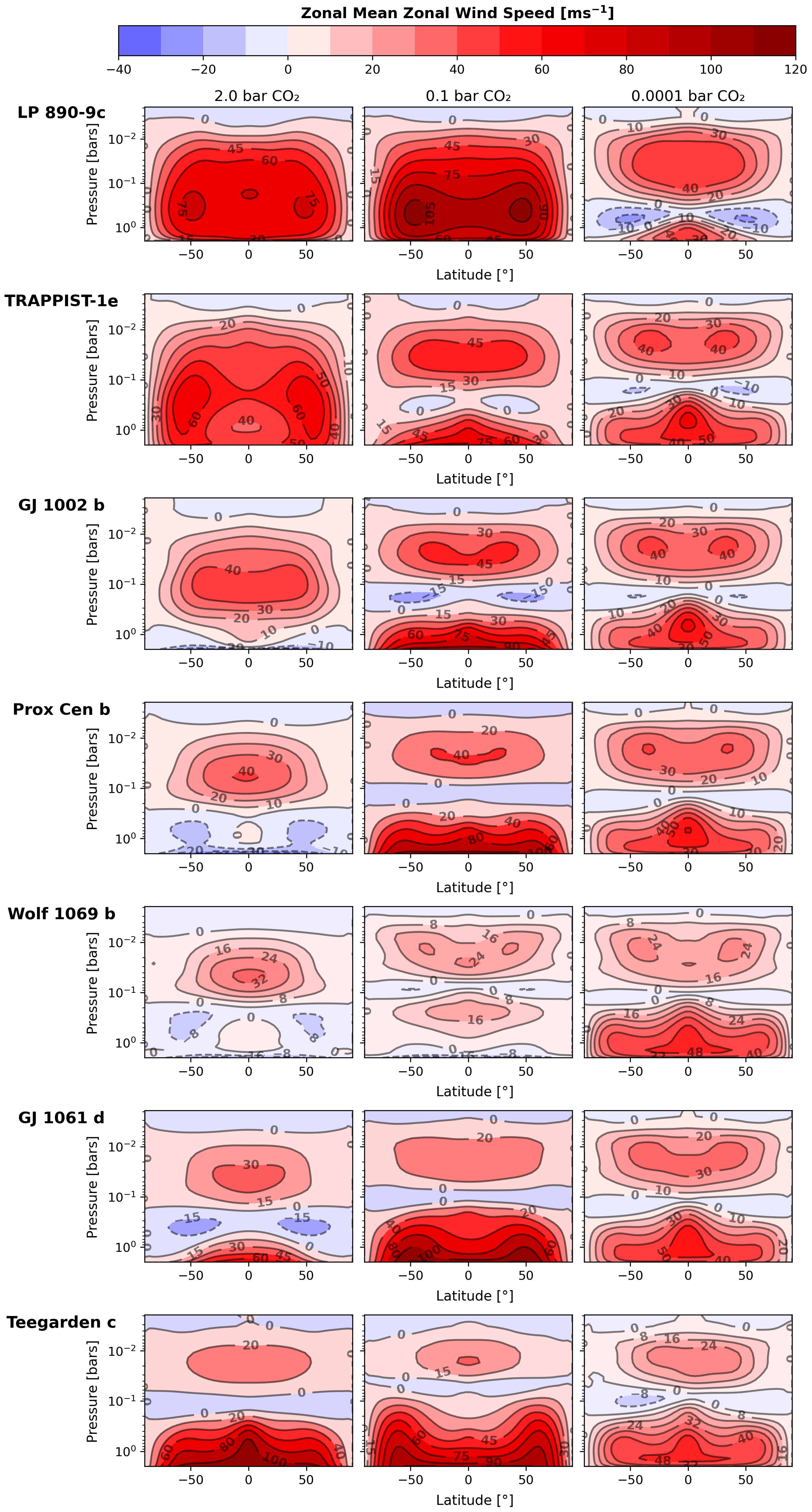}
\caption{Zonal-mean zonal wind (i.e., east-west average of the eastward wind, over all longitudes for a given latitude and pressure) from all planet cases with varying pCO$_2$. The y-axis shows pressure in bars, and the x-axis shows latitude in degrees. All plots share a colorbar. We find that all cases exhibit a superrotating jet, and typically find faster jets in cases with intermediate pCO$_2$.}
\label{fig:zonalmeanzonalwind}
\end{figure}

\Fig{fig:wind_plots} shows the corresponding near-surface zonal wind speeds (filled and open contours) and 250 mbar winds (quivers) for the 21 simulations of targets at various pCO$_2$ values. As before, all sub-plots share a colorbar to assist with inter-comparison, but the quiver magnitudes are normalized separately for each plot. In addition, \Fig{fig:zonalmeanzonalwind} shows the corresponding zonal-mean zonal wind from all simulations.
We expect that all cases exhibit dayside convergence as well as atmospheric superrotation at some pressure level due to their slow rotation given that we assume that each planet is tidally locked \citep{Showman_2013_terrestrial_review,Pierrehumbert:2019vk}. The strength of this superrotating jet increases from the dayside to nightside, with the fastest wind speeds typically found on the nightside just west of the western terminator, equatorward of the Rossby lows of the planetary-scale Matsuno-Gill pattern \citep{Hammond:2018aa}. We find that the near-surface circulation is stronger for lower values of pCO$_2$, likely due to the larger day-to-night temperature contrasts in these cases with lower global abundances of atmospheric water vapor \citep{Haqq2018}. However, we do not find a clear dependence of circulation strength on instellation, given that the effects of instellation increasing the day-to-night forcing \citep{Hammond:2020aa} and enhanced moisture leading to latent heat that reduces the day-to-night contrast \citep{Labonte:2020aa} offset one another. 
Future work is required to develop a comprehensive theory that incorporates the effects of moisture on the day-night forcing strength and jet speeds of tidally locked planets.

\subsubsection{Dynamical regimes}
\cite{Haqq2018} classified the dynamical regimes of tidally locked temperate terrestrial exoplanets into three categories depending on the ratio of both the Rhines scale and equatorial Rossby deformation radius to the planetary radius (for alternate dynamical regime definitions, see \citealp{Noda:2017aa,Wang:2018}). The equatorial Rossby deformation radius represents the typical meridional lengthscale at which near-equatorial gravity waves can propagate before being deflected by the Coriolis force, and can be expressed as
\begin{equation}
    \label{eq:Rossby}
    \lambda_{Ro} = \sqrt{\frac{R_p\sqrt{gH}}{4\Omega}}~\mathrm{.}
\end{equation}
In \Eq{eq:Rossby}, $R_p$ is the planetary radius, $g$ is the surface gravity, $H=RT/g$ is the scale height, and $\Omega$ is the rotation rate. The Rhines scale is the meridional scale at which turbulence manifests into zonal jets. The Rhines scale can be estimated as \citep{Rhines:1975aa} 
\begin{equation}
    \label{eq:Rhines}
    \lambda_{Rh} = \pi\sqrt{\frac{R_pU}{2\Omega}}~\mathrm{,}
\end{equation}
where $U$ is the characteristic zonal wind speed. 

\begin{figure*}
    \centering
    \includegraphics[width=\textwidth]{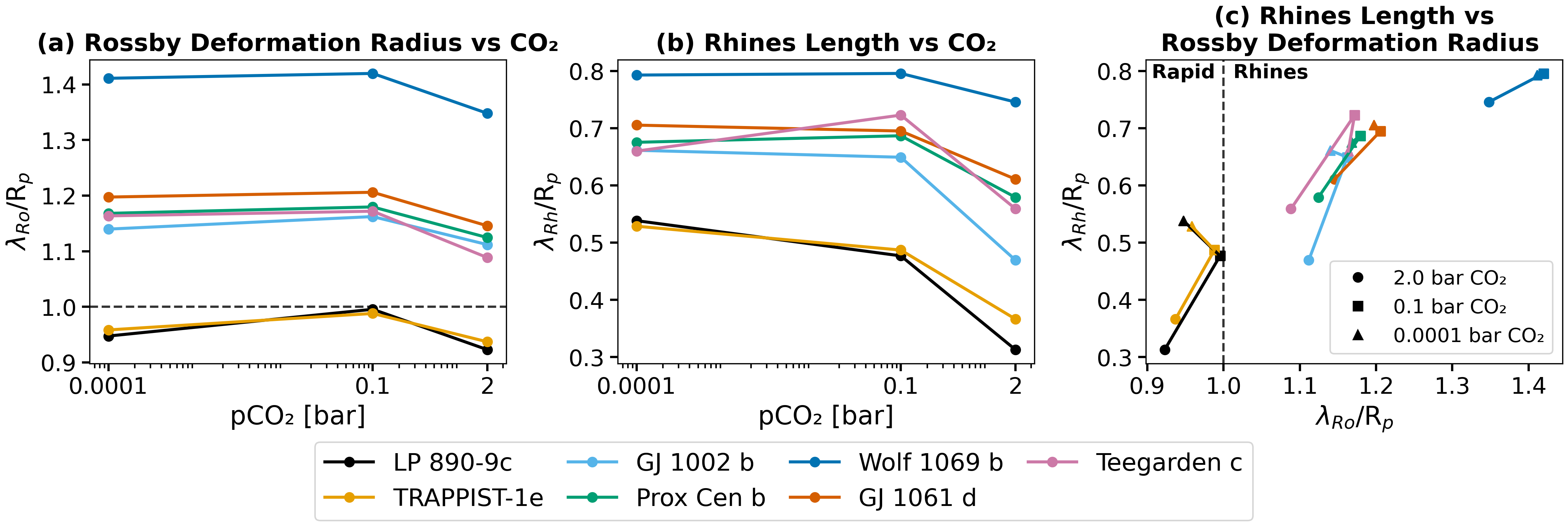}
    \caption{Plots of non-dimensional Rossby deformation radius ($\lambda_{Ro}$/$R_p$) versus pCO$_2$ (a), non-dimensional Rhines length ($\lambda_{Rh}$/$R_p$) versus pCO$_2$ (b), and non-dimensional Rhines length versus non-dimensional Rossby deformation radius (c) for each planet case. Rapid rotators are defined when $\lambda_{Rh}$/$R_p$ $<$ 1 and the $\lambda_{Ro}$/$R_p$ $<$ 1. Rhines rotators are defined when $\lambda_{Rh}$/$R_p$ $<$ 1 and $\lambda_{Ro}$/$R_p$ $>$ 1. Here, only the two highest irradiation planet cases considered (LP 890-9c and TRAPPIST-1e) are within the rapid rotator regime regardless of pCO$_2$, while all other cases are in the Rhines rotator regime.}
    \label{fig:rhines}
\end{figure*}

\cite{Haqq2018} found that temperate planets which orbit later-type host stars are more likely to be in the rapid or Rhines rotator regimes due to their shorter-period orbits and thus faster rotation rates. \Fig{fig:rhines} shows our predicted dynamical regimes from all 21 simulations of nearby rocky planets with varying pCO$_2$, showing the individual dependencies of the Rossby deformation radius and Rhines length with CO$_2$ for each planet as well as the relationship between them.
We find that all of our simulated targets lie either in the rapid or Rhines rotator regimes. This is likely because these targets were chosen for high signal-to-noise ratio with PIE, and typically orbit late-type M dwarf host stars (see Table \ref{model_params}). We predict that the two highest-irradiation targets in our sample, LP 890-9c and TRAPPIST-1e, lie in the rapid rotator regime for all pCO$_2$ considered, while all other targets lie in the Rhines rotator regime. Note that TRAPPIST-1e lies at the border of the Rhines and rapid rotator regimes, and the resulting circulation regime depends on the initial conditions and convection parameterization \citep{Sergeev:2020aa,Sergeev:2022ab}, resulting in differences in circulation regime between GCMs \citep{2022PSJ.....3..212S}. 
\subsubsection{Cloud coverage}
\begin{figure*}[h]
\begin{minipage}[c]{0.48\linewidth}
\includegraphics[width=\linewidth]{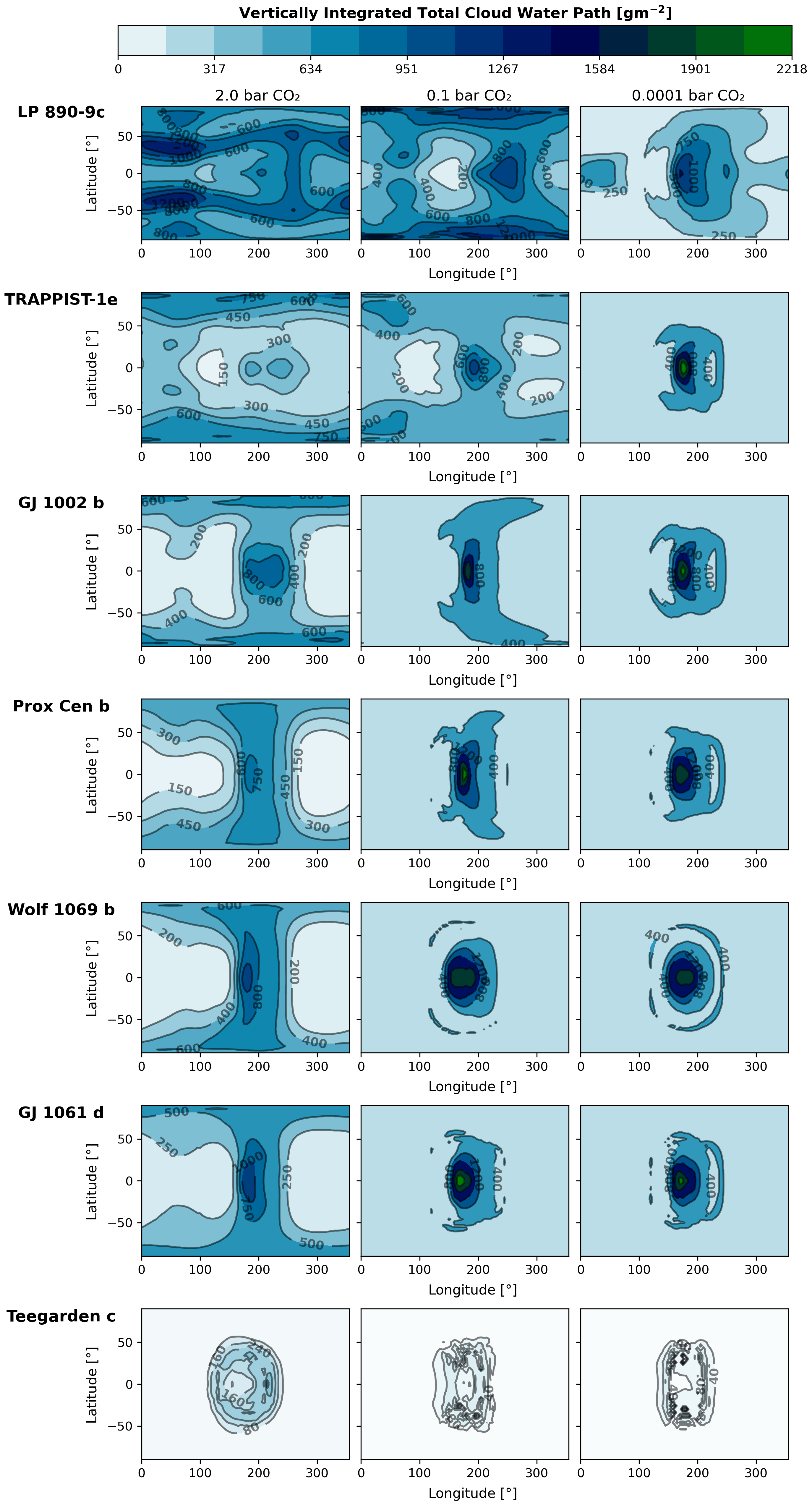}
\caption{Maps of vertically integrated total cloud water path (colors) for each GCM simulation. All cases share the same colorbar. Darker colors represent areas of higher column water mass, or more total ice and liquid cloud coverage. Hotter cases are generally cloudier as expected from Clausius-Clapeyron, and they also tend to have lower day-night cloud coverage contrast due to their lower day-to-night temperature contrasts (see \Fig{fig:st_plots}). For most cases, cloud coverage peaks just eastward of the substellar point due to advection of convectively generated clouds near the substellar point by the superrotating equatorial jet.}
\label{fig:cloudwater_plots}
\end{minipage}
\begin{minipage}[c]{0.48\linewidth}
\includegraphics[width=\linewidth]{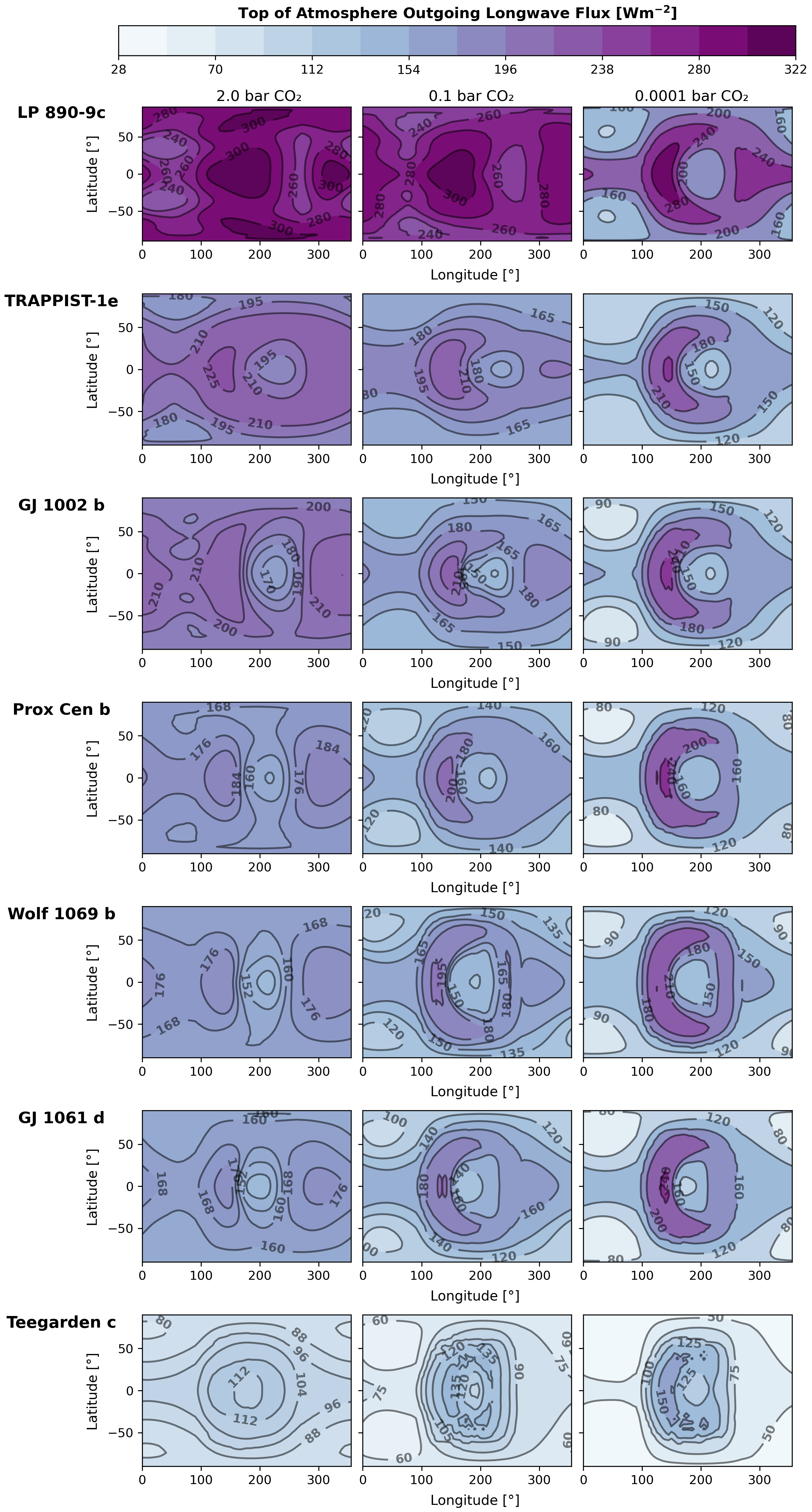}
\caption{Maps of top of atmosphere outgoing longwave flux (colors) for each GCM simulation. All cases share one color bar. Darker (purple) colors represent areas of greater infrared (longwave, thermal) emission. Cases with higher overall surface temperatures emit more overall thermal emission to space. Thermal emission to space generally peaks westward of the substellar point, since this is where cloud coverage is lowest.}
\label{fig:OLR_plots}
\end{minipage}%
\end{figure*}

Because cloud coverage controls the thermal emission of temperate rocky planets \citep{Yang:2013,Yang:2014} and is shaped by dynamics on a range of scales \citep{Sergeev:2020aa}, we next show the total vertically integrated cloud water path in \Fig{fig:cloudwater_plots}. We find that planet cases with higher instellation and higher pCO$_2$ typically have a greater global extent of the cloud water path, or a higher density of clouds. This is because cold start cases with low instellation and low pCO$_2$ only have clouds concentrated near the substellar point due to the ``eyeball'' pattern of open ocean \citep{Pierrehumbert:2011aa}. Note that if we instead used a hot start as in \cite{Turbet:2021aa,Turbet:2023aa}, the cloud distribution could be vastly different and instead comprised of nightside clouds and a cloud-free dayside due to the water vapor feedback. 

There is a slight shift of the cloud pattern eastward of the substellar point due to the predominantly eastward near-equatorial flows transporting clouds lofted near the substellar point downwind. However, we do not include ocean dynamics in these models, which could induce a greater zonal offset in peak ocean temperatures and thus cloud coverage in our moderate-to-low pCO$_2$ cases assuming an aquaplanet surface \citep{Hu:2014aa}. We find that the case of Teegarden's Star c is an outlier in the total cloud water path, given that the surface is ice-covered and the total atmospheric water vapor content is much smaller than the other six planet cases. 

\subsubsection{Bolometric top-of-atmosphere thermal emission}

We next turn to consider the impacts of our simulated climate on the top-of-atmosphere outgoing longwave flux (OLR) that can be probed via phase curves and thermal emission spectroscopy. First, we show spatial maps of the bolometric top-of-atmosphere OLR from our 21 ExoCAM GCM cases in \Fig{fig:OLR_plots}. As expected, we find that the total top-of-atmosphere OLR increases with increasing instellation, as required to maintain planetary energy balance. We further find that the pattern of top-of-atmosphere OLR is anti-correlated with the cloud water path, as expected due to the cool cloud tops reducing the thermal emission to space via the cloud greenhouse effect \citep{Seeley:2019um}. This means that the top-of-atmosphere OLR typically peaks westward of the substellar point \citep{Yang:2013}, especially in the cases with moderate and low (0.1 bar and $100~\mu\mathrm{bar}$) pCO$_2$. This is because of the superrotating eastward equatorial jet that blows the peak of the cloud water path eastward of the substellar point, leading to a deficit of clouds on the western dayside and greater surface thermal emission reaching space.

\begin{figure}
    \centering
    \includegraphics[width=\linewidth]{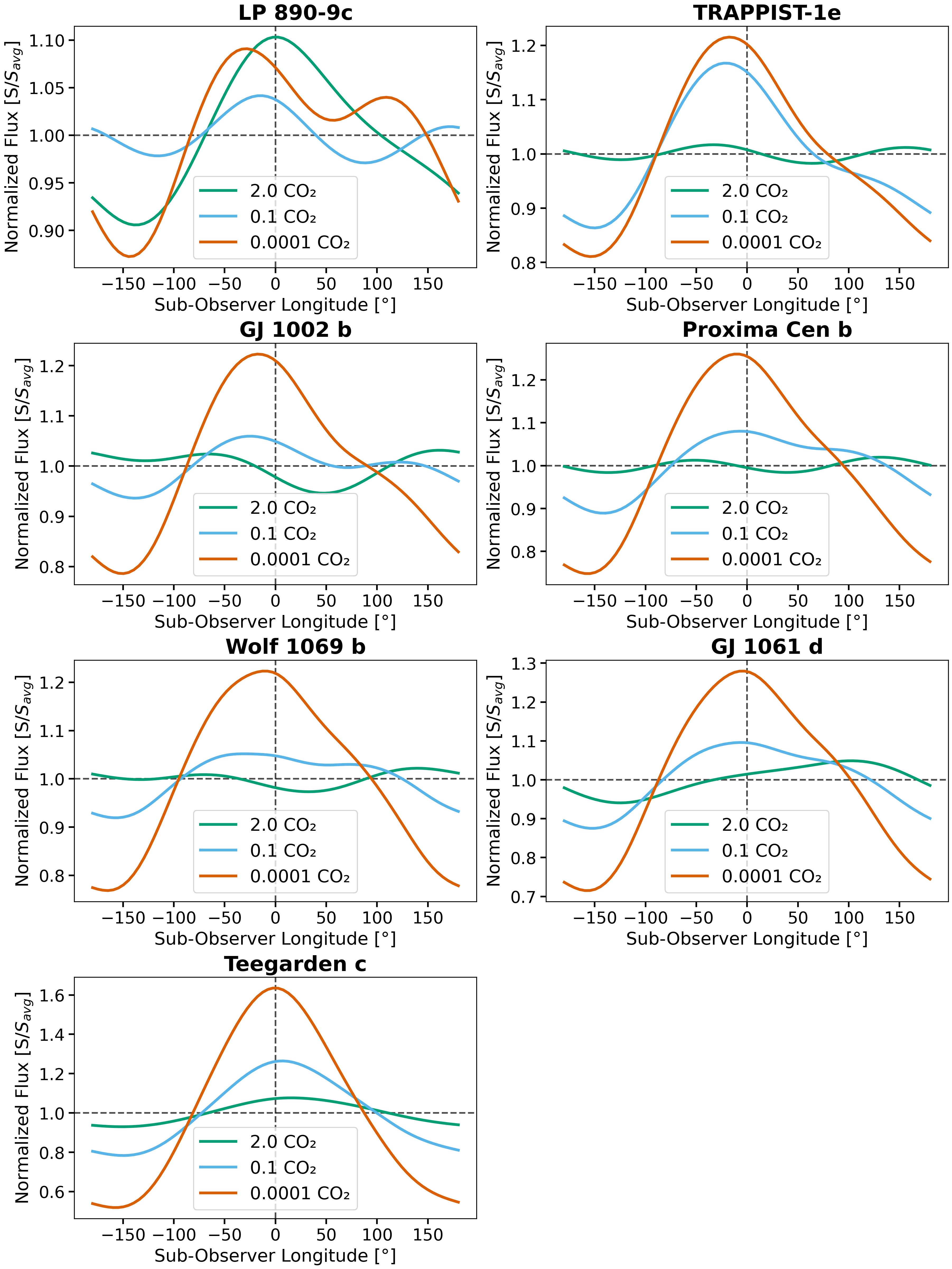}
    \caption{The normalized broadband thermal phase curve for all three pCO$_2$ cases for each planet simulated with ExoCAM. These bolometric phase curves are calculated from the ExoCAM radiative transfer output outgoing longwave radiation at the top-of-atmosphere summed over wavelength. Each phase curve is normalized to its own average value in order to compare the relative phase curve amplitudes between planets and with varying pCO$_2$. Note that each plot has a different y-axis scaling, with Teegarden's Star c having a much larger phase curve amplitude than LP 890-9c. As in \cite{Haqq2018}, we find that phase curve amplitudes increase with decreasing planetary temperature due to the limited effect of water latent heat on heat transport.}
    \label{fig:exo_phase_curves}
\end{figure}

In reality, planets are observed as point sources, with the strongly spatially dependent top-of-atmosphere OLR observed as a hemisphere average. We show such simulated orbital phase curves of bolometric thermal emission from our ExoCAM GCMs in \Fig{fig:exo_phase_curves}. Note that each panel has a different y-axis scale. We calculate phase curves as in \cite{Cowan:2008}, integrating the global outgoing longwave flux at each latitude and longitude taking into account viewing angle to obtain the total outgoing longwave flux for each hemisphere. 
Each phase curve for a given target and value of pCO$_2$ is normalized to its mean value in order to facilitate inter-comparison, and phase curves are plotted as a function of sub-observer longitude (rather than e.g., orbital phase) assuming that the hemisphere the observer sees is centered on the equator -- however, note that many of these planets are non-transiting, and thus the sub-observer latitude is a priori unknown \citep{Rauscher:2017wm}. As expected, we find that typically cases with moderate instellation and pCO$_2 \lesssim 0.1~\mathrm{bar}$ have phase curves that peak westward of the substellar point due to the eastward shift in the maximum cloud coverage \citep{Yang:2013,Komacek:2019aa-terrestrial}. However, we find that cases with high pCO$_2$ and the hottest planet case (LP 890-9c) exhibit more complex phase curves that can have multiple peaks, typically one westward of the substellar point and one on the nightside eastward of the eastern limb. Meanwhile, the coolest planet that we simulated (Teegarden's Star c) has thermal emission maxima either near the substellar point or slightly eastward of it, due to the reduced impact of cloud coverage on thermal emission in this case. As a result, Teegarden's Star c shows the largest phase amplitude given the significant difference in cloud cover and resulting OLR between the dayside and nightside. In contrast, LP 890-9c has the smallest phase curve amplitude, reflecting its relatively spatially uniform OLR due to the high moisture content leading to efficient day-to-night heat transport \citep{Haqq2018,Labonte:2020aa}.

\subsection{PSG Post-Processing}
We next turn to our results from the PSG post-processing of our 21 ExoCAM GCMs for varying targets and pCO$_2$ in order to determine the detectability of spectral features in the mid-infrared using a MIRECLE-like mission with the PIE technique.
\subsubsection{Thermal emission spectra with PIE}
\begin{figure*}
    \subfloat{
        \includegraphics[width=0.49\linewidth]{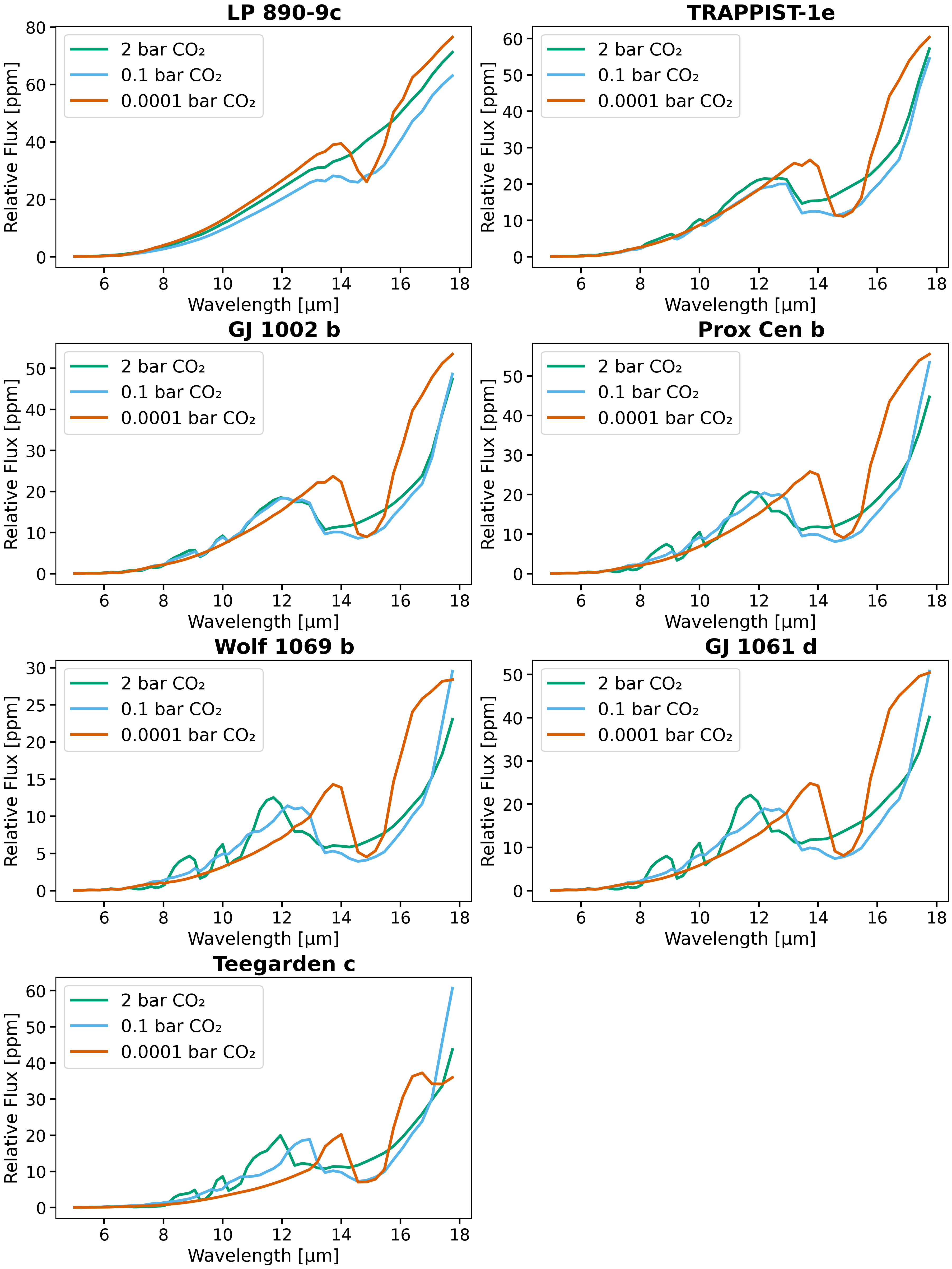}}
    \subfloat{
        \includegraphics[width=0.49\linewidth]{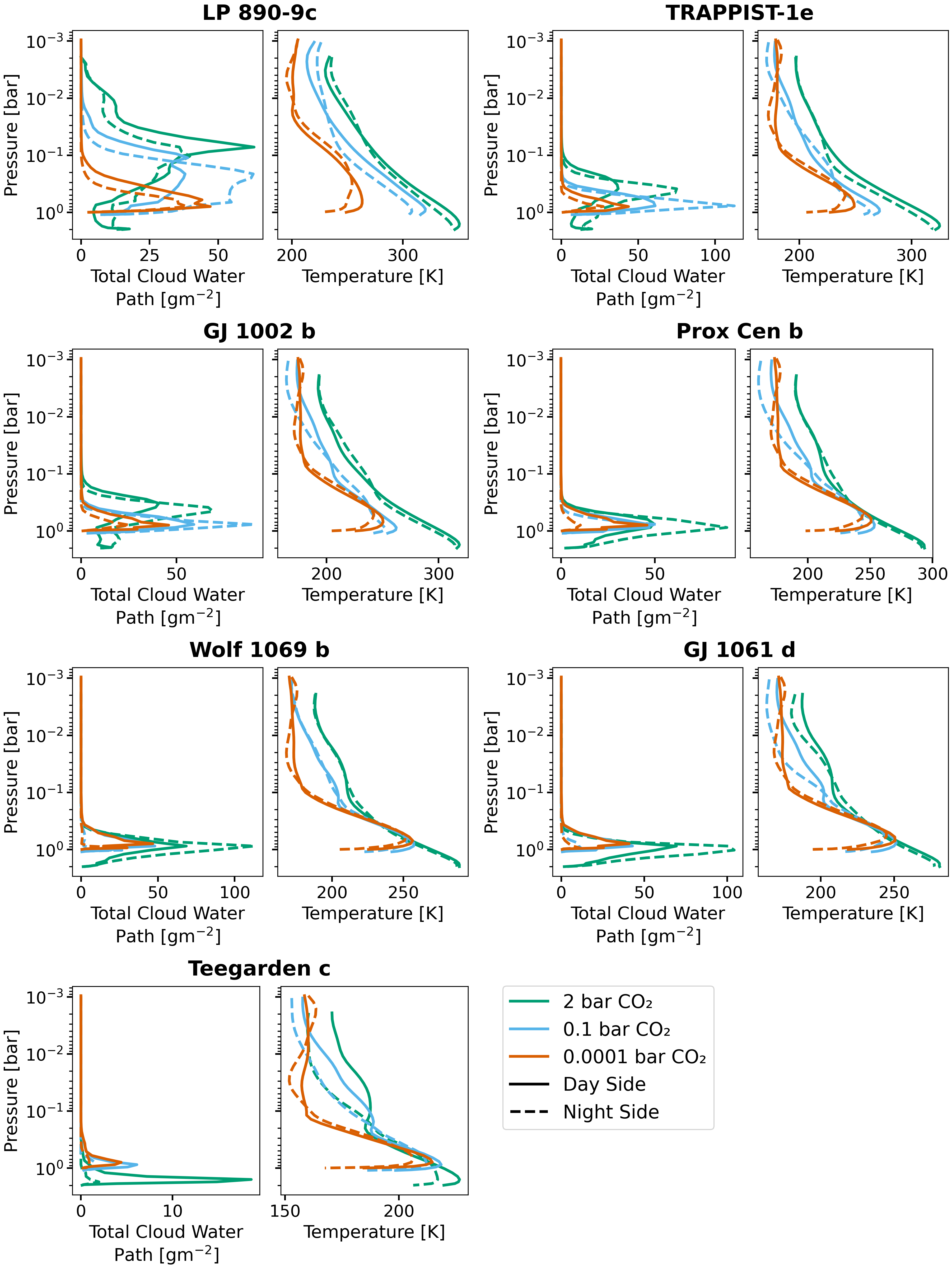}}
        
    \caption{Modeled PSG emission spectra (left-hand panels) and ExoCAM vertical profiles of cloud coverage and temperature (right-hand panels), averaged over the dayside (solid lines) and nightside (dashed lines), for each planet and pCO$_2$ case (colors). For most cases, there is a prominent absorption feature of CO$_2$ at 15 $\mu$m as well as water vapor features. We find that cases with higher instellation have reduced water and carbon dioxide feature strengths due to the presence of a greater cloud water path at low pressures.}
    \label{fig:spectra}
\end{figure*}

In Figure \ref{fig:spectra} we plot the thermal emission spectra at 270 degree phase for varying pCO$_2$ alongside vertical profiles of the total cloud water path and temperature for each target considered.

\begin{figure}
    \centering
    \includegraphics[width=\linewidth]{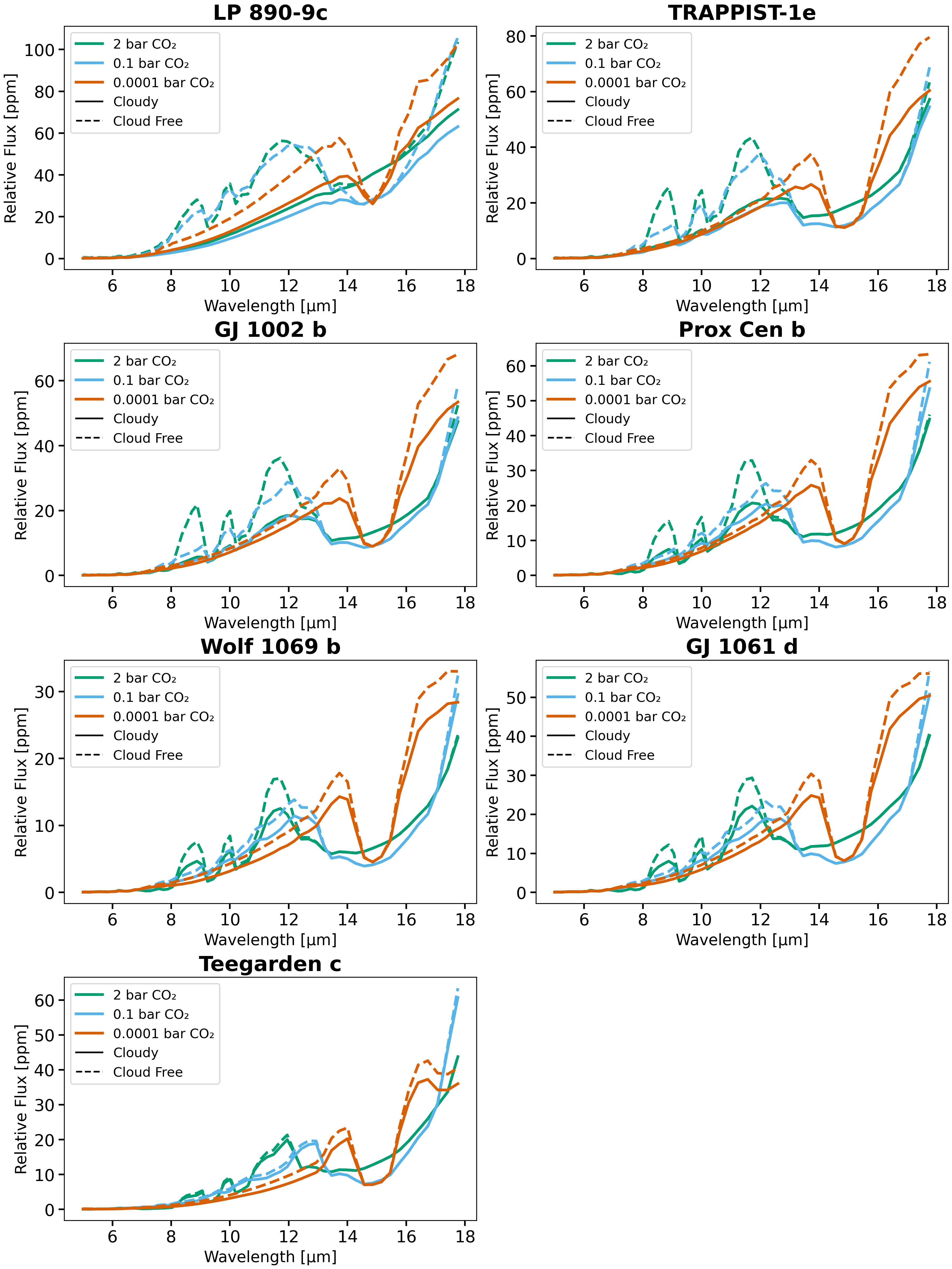}
    \caption{Modeled PSG emission cloudy (solid lines) and cloud-free (dashed lines) spectra for each planet and pCO$_2$ case (colors). Cloud-free cases represent PSG GlobES runs with no aerosol absorption. The effect of clouds on the thermal spectra is generally the highest for warm planets with a thick cloud deck. Reduced cloudiness allows for outgoing thermal emission to radiate out of the deeper atmosphere and not be absorbed and re-emitted by optically thick clouds.}
    \label{fig:spectraclouds}
\end{figure}

We find that all cases have a distinguishable CO$_2$ feature at $15~\mu\mathrm{m}$, and most have water features centered at $6.2~\mu\mathrm{m}$ and $7.3~\mu\mathrm{m}$. Figure \ref{fig:spectraclouds} shows the same set of spectra for each planet and pCO$_2$ including clouds, now compared with cloud-free counterparts simulated using PSG solely removing the impact of liquid and ice clouds but assuming that the climate is unchanged. We find that hotter cases with resulting high altitude clouds, most notably LP 890-9c, generally have weaker spectral features. 
This is because these high altitude clouds absorb and re-radiate outgoing longwave radiation originating from deeper in the atmosphere, decreasing the flux level of the continuum and thus decreasing the strength of absorption features. Similarly, we find that these cases with high-altitude (low-pressure) maxima in the cloud water path have significantly muted water vapor spectral features. Conversely, we find that cooler targets with reduced cloud cover (e.g., Teegarden's Star c) have a smaller impact of clouds on the resulting spectra.

We also find that the shape of the $15~\mu\mathrm{m}$ CO$_2$ feature is strongly dependent on pCO$_2$ itself, with higher pCO$_2$ cases having broader CO$_2$ features. This is due largely to pressure broadening with increasing pCO$_2$, and in some cases is affected by overlap between neighboring H$_2$O and CO$_2$ features in these hot and moist atmospheres with high pCO$_2$ \citep{Koll:2023aa}. In addition, the $15~\mu\mathrm{m}$ CO$_2$ feature in some of our hottest cases (e.g., LP 890-9c with pCO$_2$ = 2 bars) has a markedly different shape and lower amplitude. This is likely due to a slight thermal inversion in the upper atmosphere that is driven by water vapor absorption (given that our ExoCAM models have no O$_3$), increasing the outgoing top-of-atmosphere flux from within the CO$_2$ band. 

In reality, MIRECLE will be observing targets at distinct phases of the planet's orbit. We compute spectra at four phases: $0^\circ, 90^\circ, 180^\circ, 270^\circ$, which correspond to observing hemispheres centered on the dayside, eastern limb, nightside, and western limb, respectively. For most cases, the observed flux on the dayside is generally lower than the observed nightside flux due to high dayside cloud cover, except for the flux at the at the $15~\mu\mathrm{m}$ CO$_2$ feature in cases with 2 bars of CO$_2$, which occurs as an emission rather than absorption feature on the dayside in our hottest cases considered. Notably, this $15~\mu\mathrm{m}$ feature on the dayside of LP-890-9c is very strong due to the high amount of OLR emitted on the dayside (see Fig. \ref{fig:OLR_plots}). Conversely, the observed dayside of Teegarden's Star c has much larger relative flux because the planet is so cold and as a result there are fewer clouds to absorb and re-radiate outgoing flux from the surface.

\begin{figure*}
    \centering
    \includegraphics[width=0.8\textwidth]{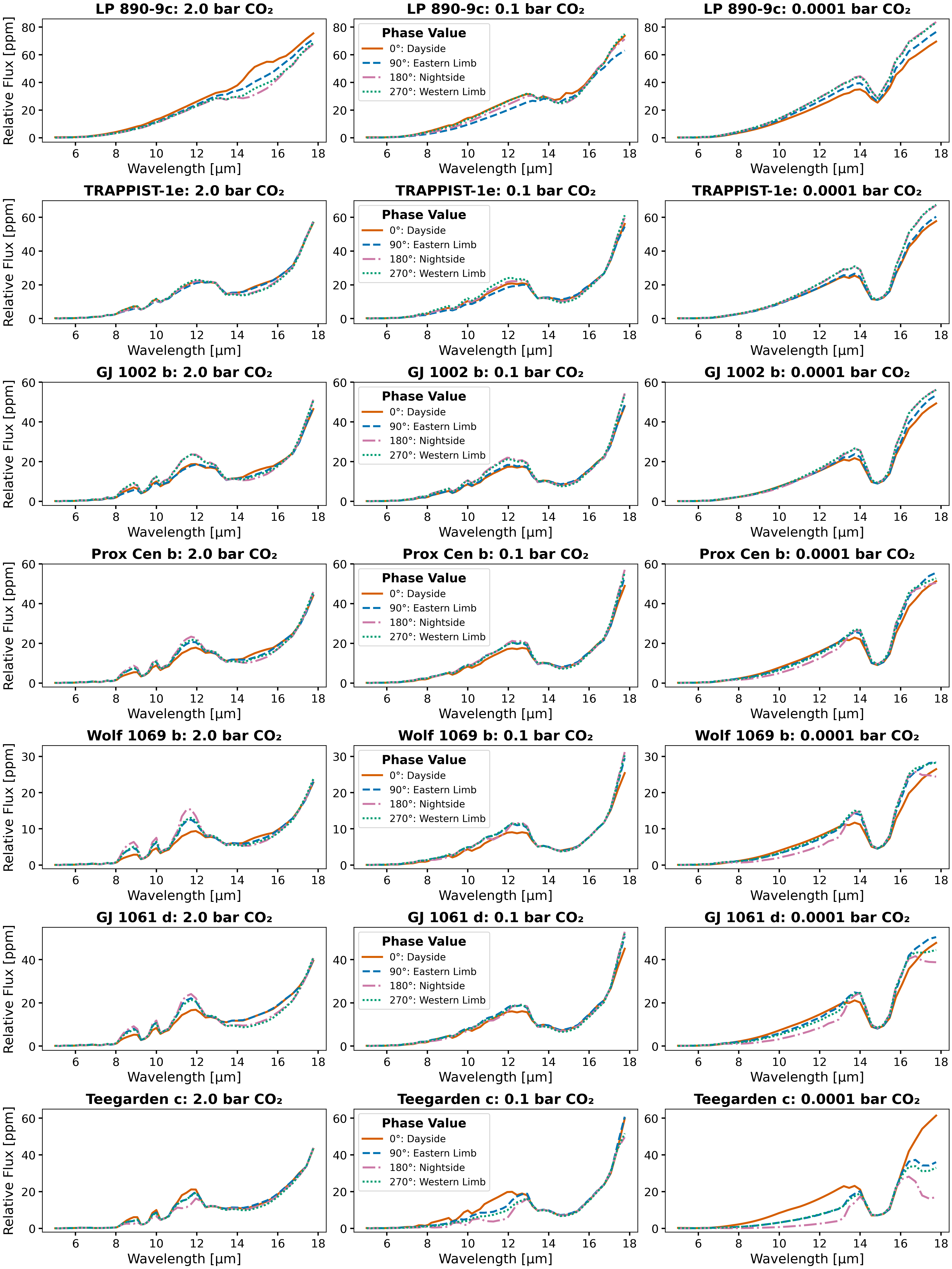}
    \caption{Modeled PSG emission spectra at four phases (colors and linestyles) for each planet and pCO$_2$ case. Especially for planets with reduced instellation (where rows are ordered with decreasing instellation from top to bottom), the water features when observing the dayside have lower in-band fluxes than when observing the nightside except for Teegarden's Star c. For the 2 bar CO$_2$ cases, the flux in the line core of the 15$\mu$m CO$_2$ feature observed on the dayside is typically higher than that on the nightside.}
    \label{fig:spectraphase}
\end{figure*}

\subsubsection{Detectability of CO$_2$ with MIRECLE}
We now use our simulated spectra shown in \Fig{fig:spectra} to quantify the detectability of CO$_2$ in nearby temperate rocky planet atmospheres. We choose CO$_2$ because it has the highest feature amplitude and because constraining the abundance of CO$_2$ on many exoplanet atmospheres could provide an inference about the feasibility of the habitable zone hypothesis \citep{Bean:2017aa,Lehmer:2020aa,Checlair:2021aa}. We calculate the molecular SNR as \citep{Lustig-Yaeger:2019aa,Rotman:2023aa}:
\begin{equation}
    \label{eq:SNR}
    \mathrm{SNR} = 
    \sqrt{\sum_\lambda \left( \frac{F_{\lambda,\mathrm{CO_2}} - F_{\lambda,\mathrm{noCO_2}}}{\sigma_\lambda} \right)^{2}} ~\mathrm{,}
\end{equation}
where $F_{\lambda,\mathrm{CO_2}}$ and $F_{\lambda,\mathrm{noCO_2}}$ are the wavelength-dependent planetary thermal fluxes in simulations with and without including CO$_2$ in the PSG post-processing and $\sigma_\lambda$ is the wavelength-dependent noise from the PSG noise calculator. 

\begin{figure}
    \centering
    \includegraphics[width=0.5\textwidth]{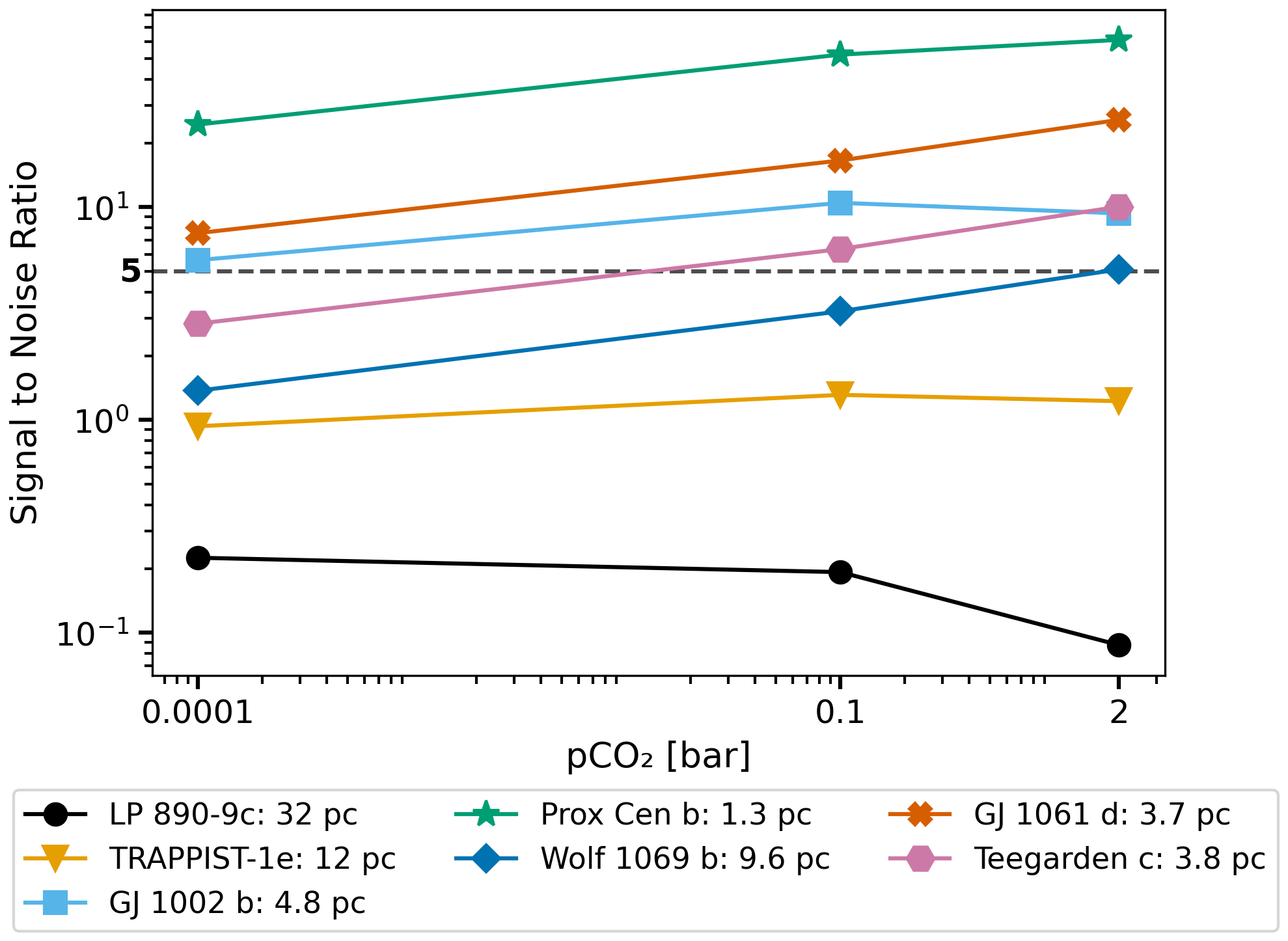}
    \caption{The calculated CO$_2$ molecular signal to noise ratio for each model case summed over all wavelengths for 30 days of observation with MIRECLE. CO$_2$ in the atmospheres of cases above the noise floor (5 ppm) are more likely to be detected by the MIRECLE telescope with our chosen instrumental parameters. In general, temperate planets within $\sim$5 pc are the best targets to potentially detect CO$_2$.}
    \label{fig:spectra2}
\end{figure}

\begin{figure}
    \centering
    \includegraphics[width=0.49\textwidth]{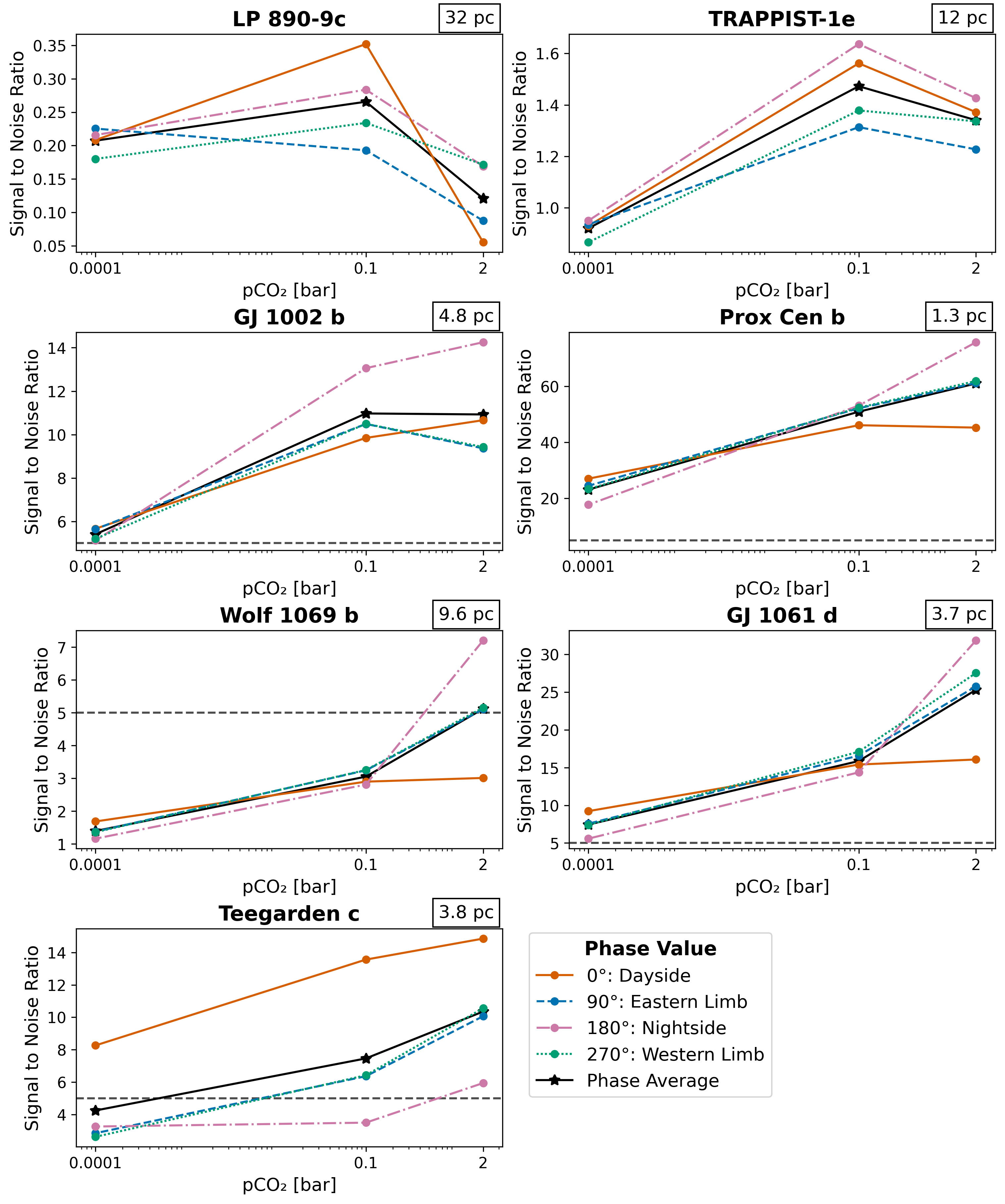}
    \caption{The calculated CO$_2$ molecular signal to noise ratio at four phase values (colors and linestyles are the same as in Figure 10) for each model case summed over all wavelengths given 30 days of observation with MIRECLE. The black line with star markers represents the SNR averaged over all phases. For some cases, the phased averaged SNR is greater than that at $270^\circ$ alone, the phase set in Figures \ref{fig:spectra} and \ref{fig:spectra2}.}
    \label{fig:snr_phase}
\end{figure}

\begin{figure}
    \centering
    \includegraphics[width=0.49\textwidth]{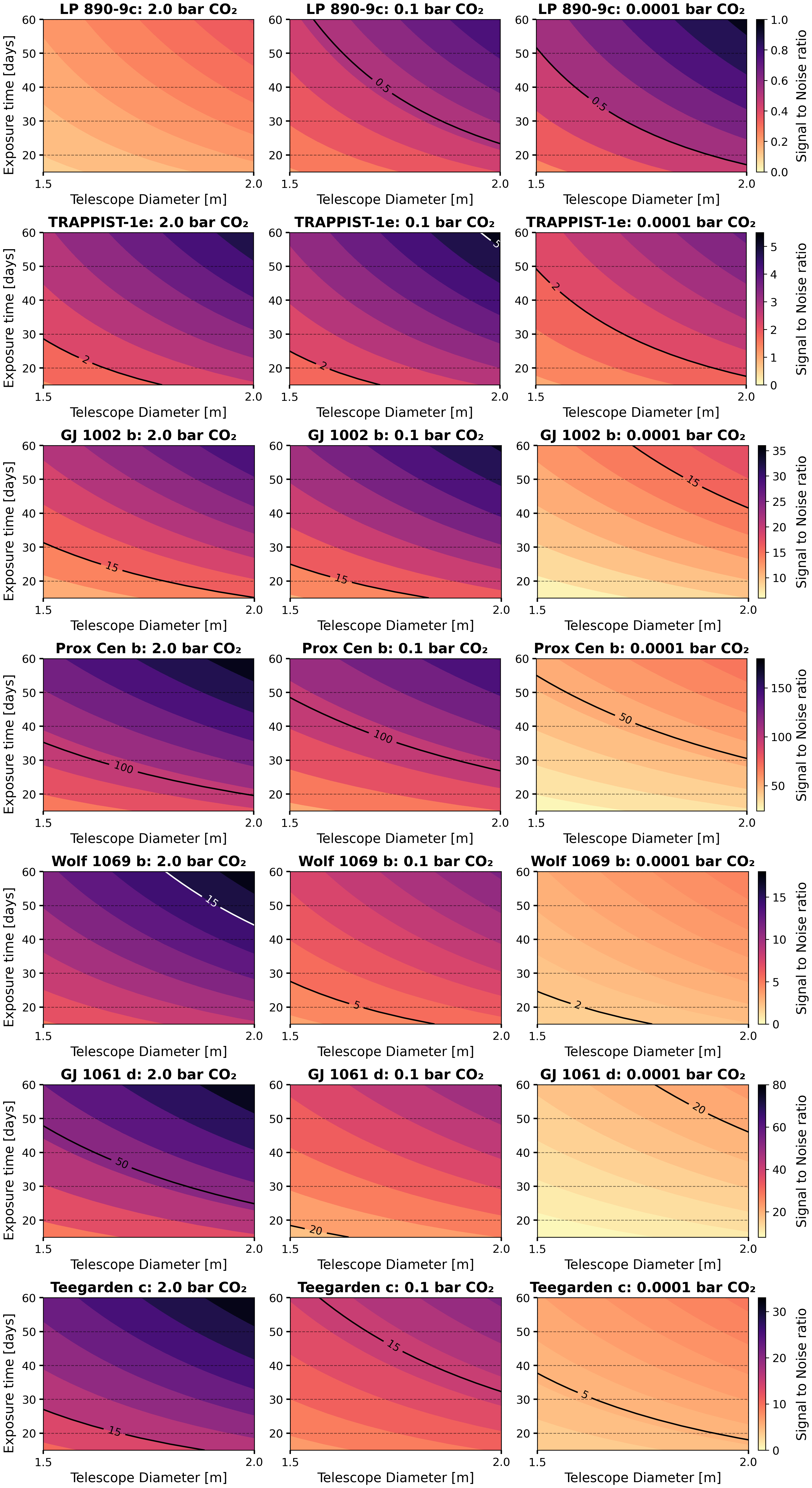}
        \caption{The calculated molecular signal to noise ratio (colors and contour lines) given varying exposure times and MIRECLE telescope diameters for each model case. Each row uses a different colorbar range, so as to compare cases with varying  CO$_2$ for a given planet to each other. We find that SNR increases with increasing exposure time and telescope diameter, as expected. With about 10 days more exposure time, the 2 bar CO$_2$ case for Wolf 1069 b would be within our 5 SNR limit for detectability. CO$_2$ will likely not be detectable on LP 890-9c within a reasonable exposure time and telescope diameter, but possibly could for TRAPPIST-1e.}
    \label{fig:snr_contours}
\end{figure}

\Fig{fig:spectraphase} shows the simulated spectra for each case at the four different phases considered, while \Fig{fig:spectra2} shows the molecular signal-to-noise ratio (SNR) for CO$_2$ for each of our targets and considered pCO$_2$ cases assuming 30 days of observation with MIRECLE. 
Figure \ref{fig:snr_phase} shows the SNR calculated at each phase. Note that the $0^\circ$ and $180^\circ$ cases are adjusted to $2^\circ$ and $182^\circ$, respectively, to account for transit and secondary eclipse given that some of our considered targets are transiting.
Finally, Figure \ref{fig:snr_contours} shows the SNR calculated for a range of exposure times and two telescope diameters. Note that 1.5m is our default telescope diameter for all previous simulations.

We find that the SNR for CO$_2$ given 30 days of observation with MIRECLE is always greater than five for three planet targets regardless of pCO$_2$: Proxima Centauri b, GJ 1061 d, and GJ 1002 b. We also find that if pCO$_2 \gtrsim 0.1~\mathrm{bar}$, CO$_2$ may be detectable on Teegarden's Star c. Clouds have little impact on the SNR as the CO$_2$ is expected to be well-mixed above the cloud deck. Phase, however, can influence detectability, particularly for cold planets (e.g., Teegarden's Star c) given their large predicted day-to-night temperature contrast, as well as hotter planets (e.g., LP 890-9c) at intermediate values of pCO$_2$ due to strong day-to-night cloud coverage variations. In general, it is worthwhile to observe a planet with PIE at multiple phases both in order to better build up signal-to-noise as well as to infer significant day-night differences that could provide additional constraints on the planetary climate.

Note that the key effect in our SNR calculation is distance from the system, as all three systems that we confidently predict would have detectable pCO$_2$ given an otherwise Earth-like atmosphere are nearby, with separations $\le 4.8~\mathrm{pc}$. The only exception to this is for very cold planets, as Teegarden's star is $\sim 1 \mathrm{pc}$ closer than GJ 1002 but the SNR of Teegarden's Star c is much lower than GJ 1002 b for pCO$_2 \lesssim 0.1~\mathrm{bar}$. 
Thus, MIRECLE is well-suited to studying the thermal emission of temperate rocky planets via PIE for sufficiently nearby targets, with Proxima Centauri b serving as a best-case scenario. 

\section{Discussion}
\label{sec:disc}
\subsection{Implications for characterization of temperate rocky planets in thermal emission}
We find from our suite of ExoCAM GCMs and PSG radiative transfer post-processing that the climates of temperate rocky exoplanets play a key role in their potential to be characterized in thermal emission with the PIE technique. Our cases with higher instellation have a higher surface temperature, leading to a greater amount of atmospheric water vapor and thus higher overall cloud coverage and a reduced day-to-night temperature contrast \citep{Yang:2013,Haqq2018,Labonte:2020aa}. 

Notably, at higher surface temperatures (due to high instellations and/or high pCO$_2$) the cloud deck moves upward to higher altitudes. These higher-altitude cloud decks absorb and re-emit thermal radiation, reducing the flux of the continuum and thus reducing the amplitude of spectral features of key molecular ro-vibrational bands. Thus, clouds can hamper detections of molecular features in thermal emission, but not to as great of an extent as transmission given the larger optical path in that case \citep{Fortney05c,Fauchez:2019aa,Suissa:2020aa}. In addition, cooler cases have a reduced amount of opaque high-altitude clouds but at the expense of a reduced thermal flux, implying a smaller signal with the PIE technique. As a result, detections of spectral features in the mid-infrared with PIE are likely most feasible for planets that are sufficiently hot to have high thermal flux while cool or dry enough to prevent the formation of opaque high-altitude cloud decks. 

The day-night variation in our simulations provides crucial insight into the climate and rotational dynamics of the observed planet. 
These flux variations can be influenced by rotation rate \citep{Haqq2018,Pierrehumbert:2019vk}, atmospheric composition, specifically greenhouse gasses \citep{Turbet:2016aa,Wolf:2017aa}, which further shape global and vertical temperature patterns, and cloud/haze coverage \citep{Yang:2013,Fauchez:2019aa}. Therefore, observations at multiple phases could provide valuable constraints on a planet's rotation rate and determine if a given planet has a stable climate or if it exhibits a run-away greenhouse state.

\subsection{Relevance to Venus and its analogs}

The simulations provided in this work have demonstrated the potential detectability of CO$_2$ at 15~$\mu$m. The presence of atmospheric CO$_2$ is expected in a variety of terrestrial planet evolution scenarios which, in extreme cases, can result in a thick CO$_2$ dominated atmosphere. The detection of such a scenario may enable the characterization of the evolutionary state of the planet as having entered a post-runaway greenhouse phase, similar to that of present day Venus. Identifying the major factors that caused Venus' climate to diverge so drastically from that of Earth is paramount for understanding the evolutionary pathways of terrestrial planets and the formation of habitable worlds \citep{kane2019d,kane2024b}. The study of exoplanets in the Venus Zone (VZ), or exoVenuses, are a complimentary pathway to learning about Venus and its history \citep{kane2014e,kane2018d,ostberg2019,kane2022b,ostberg2023a}. Indeed, the further study and understanding of the evolution of Venus’ atmosphere and its present state provides a foundation of planetary habitability on which the interpretation of exoplanet observations depends \citep{kane2021d}.

Exoplanet models are ``ground truthed'' through the analyses of atmospheric data for our sibling planet, such as those that will be provided by the coming deployment of the Deep Atmosphere Venus Investigation of Noble gases, Chemistry, and Imaging (DAVINCI), designed to accurately measure the pressure, temperature, composition, and chemistry of the Venus atmosphere all the way down to the surface \citep{garvin2022}. Current transmission spectroscopy methods, and the interpretation of those data, are limited by the relatively high atmospheric scale height of those measurements and the presence of hazes, making it difficult to effectively distinguish between Venus and Earth analog atmospheres \citep{ehrenreich2012a,barstow2016a,lustigyaeger2019b,ostberg2023c}. For example, JWST has provided the opportunity to study the atmospheres of potential exoVenuses, including TRAPPIST-1 b and c, the results of which are consistent with little to no atmosphere \citep{greene2023,zieba2023}. The emission spectra described in this work would yield important diagnostics that could significantly reduce the ambiguity between temperate and post-runaway greenhouse scenarios for nearby rocky planets in the habitable zone.

\subsection{Limitations}
\label{sec:limitations}
Our modeling framework was simplified in order to serve as an exploration of the effects of three-dimensional climate properties on the thermal emission of nearby rocky exoplanets as observed with the PIE technique, across a diverse selection of exoplanets. As a result, we made a variety of necessary simplifying assumptions regarding system and planetary properties, observation geometry, as well as atmospheric composition and variability. 

First, we assumed that all of the planet targets we studied were tidally locked to their host star, with a 1:1 spin-orbit ratio. Though this is well motivated through short spin-synchronization timescales for such close-in orbits, it is not known if these targets are indeed tidally locked. For instance, TRAPPIST-1e may or may not have a non-zero obliquity \citep{Guerrero:2023aa,Millholland:2024aa}, and in general the spin of planets in close-in compact near-resonant systems may be chaotic and lead to strongly time-dependent atmospheric circulation \citep{Chen:2023aa}. In addition, it is plausible that close-in planets without known companions are in a 3:2 rather than 1:1 spin-orbit resonance, leading to significant differences in their simulated climate states \citep{Turbet:2016aa}. 

A key limitation of this work is that the radii of non-transiting planets are unknown, and only the minimum mass is formally constrained. As a result, here we chose radii and surface gravities from the literature that assume a broadly Earth-like interior composition, but it is feasible that specific targets will have a different composition than expected based off of population-level mass-radius relationships (e.g., \citealp{Chen:2017aa}), requiring GCMs to formally consider a range of radii surface gravities. It is also feasible that the actual planet masses are significantly larger than the minimum mass, as used in this study.

Our simulated observations were simplistic in that we only considered one inclination and one planetary phase for all observations. All observations assumed a $90^\circ$ inclination, but if the system is more inclined, the planetary thermal emission signal will decrease due to a greater hemispheric contribution from the colder poles. As a result, we anticipate that the predictions presented here are upper limits for the detectability of CO$_2$ on these targets with PIE.
In addition, PSG uses \citet{Kurucz_2005} stellar templates for all spectra calculations. However, in reality M-dwarf stars are variable and true spectral models can complicate interpretation of planetary spectral features, especially those that also appear in the stellar photosphere. 

We neglected the potential time-variability of the atmospheric circulation of these planets in our post-processing of GCM simulations. This variability in cloud and water vapor distributions have been shown to have a $\sim$10 ppm-level impact on transmission spectra of TRAPPIST-1e \citep{May:2021aa,Fauchez_2022, Rotman:2023aa}, with ExoCAM showing the highest amplitude and longest periodicity of variability out of the four GCMs included. We neglected ocean dynamics in our ExoCAM GCM simulations, which would affect the surface temperature distribution and potentially impact resulting cloud patterns and top-of-atmosphere OLR \citep{Hu:2014aa,Salazar:2020aa,Batra:2024aa}. We also neglected the potential impact of photochemistry on atmospheric composition \citep{Chen:2021wu,Braam:2023aa}, especially any time-dependent photochemistry and atmospheric loss due to the high flaring activities of these late-type stars \citep{Amaral:2022aa,Fromont:2024aa,Ealy:2024aa}. Future work is required to quantify whether time-dependent atmospheric chemistry could be detectable in thermal emission with MIRECLE and/or LIFE. 

\subsection{Future work}
Our preliminary study considering seven nearby targets with a terrestrial planet GCM is promising in that we find that key molecular features may be detectable with PIE using a MIRECLE-like observatory. However, the limitations of this work described above motivate significant follow-up work to explore the broad range of potential planetary properties, atmospheric composition, and observing geometries. 

One key constraint regarding the atmospheric composition is that it does not take into account photochemistry, especially haze formation in methane-rich atmospheres \citep{2016AsBio..16..873A,2017ApJ...836...49A}. Recent work has shown that haze formation due to methane photodissociation on TRAPPIST-1e can significantly impact planetary climate \citep{Mak:2024aa}. Notably, \cite{Mak:2024aa} found that hazes could either cool or warm the underlying atmosphere depending on the ratio of pCO$_2$ to pCH$_4$. Future work is needed to couple a 3D GCM and a 1D photochemistry model for simulations of nearby non-transiting rocky planet targets in order to make predictions for the resulting methane and haze profiles and their impact on climate and observable properties. 

 At the time of this study, ExoCAM does not natively include oxygen and ozone, yet they have features in the mid-infrared \citep{Fauchez:2020aa}. Separate work is already underway to formally include absorption oxygen species and associate collision-induced-absorption (Dietrick et al., in prep.). In addition, the 3D ozone distribution on tidally locked planets is dynamic, with ozone preferentially accumulated in the cool Rossby gyres on the nightside and limb \citep{Braam:2023aa}. Thus, future work is required to include both methane and ozone photochemistry in ExoCAM in order to make detailed predictions for biosignature detection in the atmospheres of nearby temperate rocky exoplanets. 

\section{Conclusions}
\label{sec:conc}
In this work, we conducted a large suite of 21 GCM simulations of prime nearby habitable zone rocky planet candidates with the ExoCAM GCM for varying assumed atmospheric compositions, varying the partial pressure of carbon dioxide from $100~\mu\mathrm{bar}$ to 2 bars. We then post-processed these GCM simulations with the Planetary Spectrum Generator to make predictions for potential observational characterization of these targets with the PIE technique via thermal emission spectra and phase curves. We summarize our key findings as follows.
\begin{enumerate}
    \item Out of the seven nearby rocky planet targets we simulated, six can have habitable surface conditions over the range of CO$_2$ partial pressure that we considered. The only exception is Teegarden's Star c, which has a fully ice-covered surface even at a CO$_2$ partial pressure of 2 bar. We further find as expected that the climate, circulation, and temperature patterns of each target is strongly dependent on the assumed CO$_2$ partial pressure, with higher CO$_2$ abundances leading to lower day-to-night temperature contrasts and weaker wind speeds. 
    \item Our ExoCAM simulations predict extensive water cloud coverage on all six of the targets that have habitable surface conditions. This cloud coverage transitions from being largely confined near the substellar point at low CO$_2$ abundances to extending from pole to pole on the dayside at higher CO$_2$ abundance, and even on the nightside for our high instellation planet cases. As expected we find that this cloud coverage is anti-correlated with the top-of-atmosphere outgoing longwave flux, leading to a westward shift in the peak of thermal emission due to the eastward shift of the cloud maximum due to advection by the superrotating jet. 
    \item We find that clouds control the phase-dependent thermal emission of our habitable rocky planet targets. Their phase curves peak at sub-observer longitudes centered westward of the substellar point due to the decreased cloud coverage there. In addition, the 15 $\mu\mathrm{m}$ CO$_2$ feature amplitude is largest in cases with low CO$_2$ abundances due to the reduced high cloud cover. Conversely, water vapor features are stronger in cases with higher CO$_2$ abundances due to the enhanced amount of water vapor in the atmosphere.
    \item We predict that the PIE technique can enable the detection of carbon dioxide in the atmospheres of nearby habitable zone targets. Specifically, we expect that CO$_2$ is detectable for Proxima Cenaturi b, GJ 1061 d, and GJ 1002 b with less than 30 days of observation with a MIRECLE-like observatory for any of the CO$_2$ partial pressures we considered. In addition, we predict that CO$_2$ would be detectable on Teegarden's Star c 
    in the same conditions for sufficiently high CO$_2$ partial pressures. Future work is needed to determine the detectability of the range of possible habitability indicators and biosignatures in thermal emission for nearby non-transiting rocky planets with PIE.
\end{enumerate}

\begin{acknowledgments}
We thank the referee for their thoughtful report, which improved this work. We thank Kevin Stevenson for helpful discussions regarding PIE. We also thank Stephanie Olson for her support and guidance during the final stages of this project. We acknowledge support from NASA Habitable Worlds Program grant number 80NSSC24K0215. This material is based upon work performed as part of the CHAMPs (Consortium on Habitability and Atmospheres of M-dwarf Planets) team, supported by the National Aeronautics and Space Administration (NASA) under Grant Nos. 80NSSC21K0905 and 80NSSC23K1399 issued through the Interdisciplinary Consortia for Astrobiology Research (ICAR) program. VK and RK were supported by the GSFC Sellers Exoplanet Environments Collaboration (SEEC) which is supported by the NASA Planetary Science Division's Internal Scientist Funding Model, and the Exoplanets Spectroscopy Technologies (ExoSpec), which is a part of the NASA Astrophysics Science Division’s Internal Scientist Funding Model.
The authors thank the NCCS Discover and University of Maryland Zaratan supercomputing resources (\url{http://hpcc.umd.edu}) that enabled this research.
\end{acknowledgments}

\appendix
\section{Global-mean GCM output}
\label{ref:app}
Table \ref{tab:output} shows the global mean simulation output for surface temperature, day-night surface temperature contrast, cloud fraction, shortwave albedo, and top of atmosphere upward longwave flux from each planetary case considered in this work. 

\begin{table*}[h!]
\centering
\begin{tabular}{Y{2.25cm} |Y{1.5cm}|Y{2.25cm}|Y{2.25cm}|Y{2.25cm}|Y{2.25cm}|Y{2.25cm}}
\hline\hline
\textbf{Planet Name} & \textbf{pCO$_2$ [bar]} & \textbf{Global Mean T [K]} & \textbf{Day-Night $\Delta$T [K]} & \textbf{Global Cloud Fraction} & \textbf{Global Shortwave Albedo} & \textbf{Global TOA Upward LW [Wm$^{-2}$]} \\
\hline\hline
\multirow{3}{2cm}{\textbf{LP 890-9c}} & 2 & 345 & 3 & 0.55 & 0.58 & 489 \\
                                      & 0.1 & 313 & 11 & 0.80 & 0.58 & 369 \\
                                      & 0.0001 & 244 & 41 & 0.46 & 0.61 & 213 \\ \hline
\multirow{3}{2cm}{\textbf{TRAPPIST-1e}} & 2 & 320 & 6 & 0.56 & 0.58 & 327 \\
                                        & 0.1 & 265 & 18 & 0.49 & 0.59 & 209 \\
                                        & 0.0001 & 223 & 37 & 0.33 & 0.70 & 152 \\\hline
\multirow{3}{2cm}{\textbf{GJ 1002 b}} & 2 & 315 & 7 & 0.52 & 0.58 & 310 \\
                                      & 0.1 & 252 & 23 & 0.35 & 0.63 & 190 \\
                                      & 0.0001 & 221 & 43 & 0.33 & 0.69 & 151 \\ \hline
\multirow{3}{2cm}{\textbf{Proxima Centauri b}} & 2 & 293 & 10 & 0.59 & 0.58 & 238 \\
                                               & 0.1 & 237 & 33 & 0.28 & 0.66 & 169 \\
                                               & 0.0001 & 217 & 47 & 0.31 & 0.68 & 143 \\ \hline
\multirow{3}{2cm}{\textbf{Wolf 1069 b}} & 2 & 289 & 11 & 0.53 & 0.58 & 227 \\
                                        & 0.1 & 238 & 35 & 0.23 & 0.63 & 170 \\
                                        & 0.0001 & 220 & 48 & 0.31 & 0.65 & 149 \\ \hline
\multirow{3}{2cm}{\textbf{GJ 1061 d}} & 2 & 280 & 15 & 0.43 & 0.52 & 214 \\
                                      & 0.1 & 230 & 38 & 0.26 & 0.66 & 157 \\
                                      & 0.0001 & 213 & 50 & 0.29 & 0.68 & 135 \\ \hline
\multirow{3}{2cm}{\textbf{Teegarden's Star c}} & 2 & 217 & 28 & 0.24 & 0.76 & 108 \\
                                               & 0.1 & 195 & 47 & 0.27 & 0.77 & 90 \\
                                               & 0.0001 & 180 & 47 & 0.27 & 0.77 & 73 \\ \hline
\end{tabular}
\caption{Key values for each model case averaged globally over latitude and longitude, as well as time-averaged over the last 10 years of GCM output.}
\label{tab:output}
\end{table*}

\bibliography{sources, References_terrestrial, references, venus}
\end{document}